\documentclass[
 reprint,
 amsmath,amssymb,
 aps,
 prapplied
]{revtex4-2}

\usepackage{graphicx}
\usepackage{dcolumn}
\usepackage{bm}
\usepackage[utf8]{inputenc}
\usepackage{upgreek}
\usepackage{soul}
\usepackage{graphicx,color}
\usepackage{xcolor}

\newcommand{\Lc}{$L_c^{\text{iso}}$}
\newcommand{\Lcbir}{$L_c^{\text{bir}}(\varphi,\vartheta,\gamma)_n$}

\begin{document}

\preprint{APS/123-QED}

\title{Modeling of Random Quasi-Phase-Matching in Birefringent Disordered Media}

\author{Jolanda S. Müller}
\author{Andrea Morandi}
\author{Rachel Grange}
\author{Romolo Savo}
\email{savor@phys.ethz.ch}

\affiliation{Optical Nanomaterial Group, Institute for Quantum Electronics, Department of Physics, ETH Zurich, Zurich, Switzerland}

\date{\today}

\begin{abstract}
    We provide a vectorial model to simulate second-harmonic generation (SHG) in birefringent, transparent media with an arbitrary configuration of nonlinear ($\chi^{(2)}$) crystalline grains. We apply this model on disordered assemblies of LiNbO$_3$ and BaTiO$_3$ grains to identify the influence of the birefringence on the random quasi-phase-matching process.
    We show that in monodispersed assemblies, the birefringence relaxes the grain size dependence of the SHG efficiency. In polydispersed assemblies with sufficiently large grains, we find that the birefringence introduces an SHG efficiency enhancement of up to 54\% compared to isotropic reference crystals, which is grain size independent. This enhancement increases linearly with the grain size, if the birefringent grains can be phase matched. These two different scaling behaviours are used in Kurtz and Perry's powder-technique to identify the phase-matchability of a material.
    We show on the example of LiNbO$_3$ and ADP that this technique cannot be applied when the grains get smaller than the coherence length, because the SHG scaling with the grain size becomes material specific.
\end{abstract}

\maketitle
   
    Optical frequency conversion through phase matching in birefringent nonlinear ($\chi^{(2)}$) crystals has enabled a variety of applications in laser technology~\cite{boyd2008nonlinear,Garmire:13} and plays a key role in quantum source schemes~\cite{KwiatZeilinger1995_PRL_PhotonPairs}. The birefringence provides an orientation- and polarization-dependent refractive index, which is used to compensate for the phase lag accumulated by the mixing waves. This avoids the destructive interference of the generated waves, which starts to occur at lengths larger than the coherence length in any not phase-matched configuration. Random quasi-phase-matching (RQPM) is an alternative approach for removing destructive interference in optical frequency conversion~\cite{skipetrov2004RQPM,baudrier2004random}. It relies on the disordered distribution of the $\chi^{(2)}$-domains of certain media, such as polycrystalline materials~\cite{baudrier2004random,fischer2006broadband,bravo2010optical}, scattering powders~\cite{Kurtz1968/doi:10.1063/1.1656857,deBoer1993SHGcorr,faez2009SHGdiff,makeev2003second} and bottom-up assembled photonic structures~\cite{Molina2008Strontium,savo2020broadband}. Under a pump excitation the domains, or \textit{grains}, generate nonlinear waves with random phases and amplitudes. This leads to the cancellation of the interference terms and to the linear accumulation of the generated power with the number of domains. While RQPM is generally less efficient than phase matching, it provides numerous advantages, such as an ultra-broad acceptance bandwidth (tunability of the pump over hundreds of nanometers), the use of low-cost materials, and relaxed constraints on the polarization and the angle of incidence of the pump. These features make RQPM attractive for a variety of applications, such as  optical-parametric-oscillators~\cite{ru2017OPOrandom}, ultrafast mid-infrared lasers~\cite{vasilyev10.1117/12.2223828}, and ultrafast autocorrelators~\cite{Fischer2007Monitoring}. The interest on RQPM has been largely focused on isotropic (i.e. non-birefringent) crystals such as ZnSe, both experimentally and theoretically~\cite{baudrier2004random,vidal2006generation,Kawamori_PhysRevApplied2019_RQPM}---one reason for this is that RQPM enables taking advantage of the large nonlinearity of these isotropic crystals, even if they are non-phase-matchable.
    First modeling of RQPM relied on the scalar approximation of both the optical field and the $\chi^{(2)}$-susceptibiliy of the grains~\cite{vidal2006generation,chen2019non}. Very recently, more comprehensive vectorial models have been developed to study the nonlinear speckle statistics~\cite{Kawamori_PhysRevApplied2019_RQPM} and the supercontinuum generation from isotropic disordered  polycrystals~\cite{Gu2020Simulation}. SHG in birefringent disordered materials has only been modeled by averaging over the single grain generation, without considering the effects of the propagation through the birefringent medium on the polarization states~\cite{Trull2007Second,Aramburu2013Second}. Up to now, no comprehensive model that explicitly accounts for the birefringence in RQPM has been presented.
    
    Here, we introduce a vectorial model that considers the full three dimensional rotation (i.e. full $\chi^{(2)}$-tensor) as well as the birefringence of each grain. The generation and propagation of the waves is calculated from grain-to-grain, considering the phases of the ordinary and extraordinary beam components separately. This is done within the approximation of no scattering at the grain-to-grain boundaries. We explore the effects of the birefringence on RQPM by calculating the second-harmonic generation (SHG) in disordered assemblies of lithium niobate (LiNbO$_3$) and barium titanate (BaTiO$_3$).
    We find that in general the birefringence is beneficial for RQPM. In monodispersed  assemblies, the birefringence relaxes the grain size dependence of the SHG efficiency. In polydispersed assemblies, it introduces an efficiency enhancement of up to 54\% compared to an isotropic reference material. Thanks to the birefringence, our model can simulate grains in a phase-matchable regime, in which the SHG enhancement increases linearly with the grain size. This allows us to draw a comparison to the powder-characterization technique of Kurtz and Perry~\cite{Kurtz1968/doi:10.1063/1.1656857,Aramburu2013Second}, which is widely applied to identify the phase-matchability of a crystalline material.
    Contrary to the established behaviour at large grain sizes, we find that the SHG scaling with the grain size cannot unequivocally identify the phase-matchability of a crystal when the grain size is smaller than the coherence length. 
    %
    %
    \section{Model}
\begin{figure}[t!]
        \centering
        \includegraphics[scale = 1]{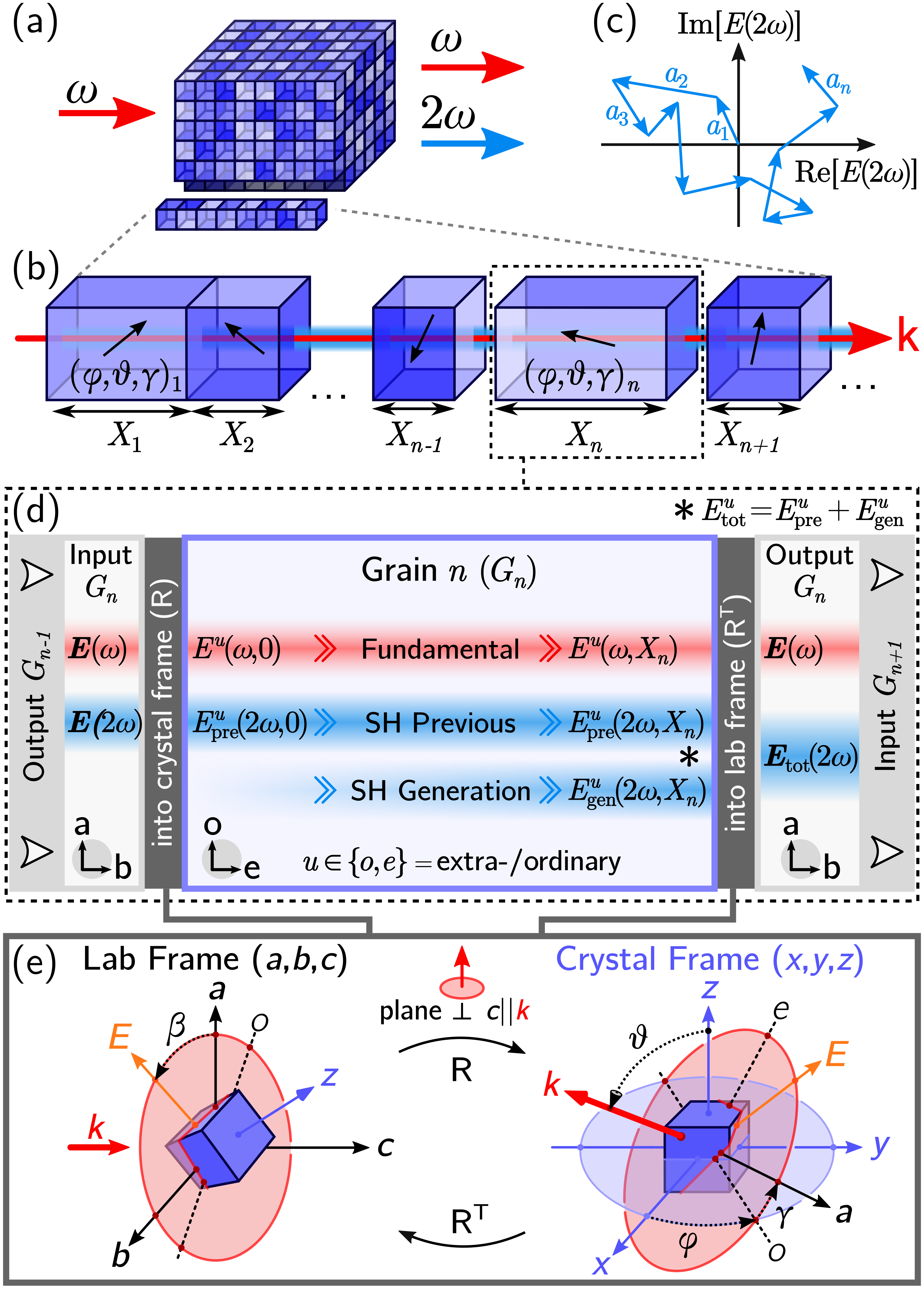}
        \caption{(a)~Sketch of a three dimensional disordered assembly of second-order nonlinear grains, consisting of multiple parallel one dimensional sticks. (b)~Each stick contains grains of varying size $X_n$ and orientation $(\varphi,\vartheta,\gamma)_n$, represented by the variation in colour. (c)~Phasor ($a_n$) representation of the interference between the second-harmonic waves generated by the grains within the stick. Each stick corresponds to a single random walk in the SHG complex plane, in which the step length is the amplitude of the second-harmonic field. (d)~Schematics of the process to propagate, generate and interfere the beams. The input (left) is transformed into the crystal frame, propagated, summed with the generated SHG, and transformed back into the lab frame. (e)~Depiction of the angles between the crystal frame ($\bm{x}$,$\bm{y}$,$\bm{z}$) and the lab frame ($\bm{a}$,$\bm{b}$,$\bm{c}$) with the transformation matrix $R$. $\beta$ is the polarisation angle of the input beam in case of a linearly polarized beam ($\phi_a = \phi_b$). The ordinary/extraordinary axes are indicated with $o$ and $e$.}
        \label{Figure_1}
    \end{figure}

    The presented model considers a disordered three dimensional cuboid structure, consisting of quadratic ($\chi^{(2)}\neq 0$), birefringent, crystalline grains, as depicted in Fig.~\ref{Figure_1}a. Similarly to other models~\cite{vidal2006generation,Kawamori_PhysRevApplied2019_RQPM}, the three dimensional problem is reduced to a one dimensional problem by breaking down the cuboid into multiple parallel one dimensional sticks. The SHG of each individual stick is calculated separately. Fig.~\ref{Figure_1}b shows one disorder configuration of such a stick, in which the grains have random sizes and orientations. The size $X_n$ is randomly chosen from a Gaussian distribution with an average grain size $\bar{X}$ and a polydispersity $\sigma$. The random orientation is defined through the rotation between the lab frame ($\bm{a}$,$\bm{b}$,$\bm{c}$) and the crystal frame of each grain ($\bm{x}$,$\bm{y}$,$\bm{z}$), with the Euler angles $(\varphi, \vartheta, \gamma)_n$, as shown in Fig.~\ref{Figure_1}e. For this, the lab frame axes are defined as the $\bm{c}$-axis parallel to the propagation direction $\bm{k}$ and the two orthogonal directions $\bm{a}$ and $\bm{b}$, while the $\bm{z}$ axis is the optic axis of the crystal. For a uniform distribution of crystal orientations, we choose the rotation around $\bm{z}$ as $\varphi\in[0,2\pi]$, the angle between $\bm{k}$ and $\bm{z}$ as $\vartheta=\arccos(u)$ with $u\in[-1,1]$~\cite{Kawamori_PhysRevApplied2019_RQPM}, and the rotation around $\bm{k}$ as $\gamma\in[0,2\pi]$. The refractive indices of the ordinary ($o$) and extraordinary ($e$) axes are given by $n_o$ and $n_e(\vartheta) = \big(\sin(\vartheta)^2/(\bar{n}_e)^2+\cos(\vartheta)^2/n_o^2\big)^{-1/2}$, with $\bar{n}_e$ the extraordinary refractive index at $\vartheta = \pi/2$~\cite{boyd2008nonlinear}. A visualization of the calculation process in one single grain is shown in Fig.~\ref{Figure_1}d. The incoming beam (pump) is defined as the plane wave (red)
    $$\bm{E}(\omega) = (\hat{\text{\bf{e}}}^{a}e^{i\phi_a}\cos\beta+\hat{\text{\bf{e}}}^{b}e^{i\phi_b}\sin\beta) E_{\omega} e^{i(\bm{k}\bm{c} - \omega t)}$$
    with amplitude $E_{\omega}$, frequency $\omega$, wavevector $\bm{k}$, the starting phases $\phi_a$ and $\phi_b$, and a polarisation angle $\beta$ in the lab frame. The vectors $\hat{\text{\bf{e}}}^{a}$ and $\hat{\text{\bf{e}}}^{b}$ are the unit vectors along the $\bm{a}$ and $\bm{b}$ axes. At the beginning of the grain, the electric fields in the lab frame $\bm{E}_{\text{lab}}$,  are transformed into the reference frame of the respective crystal $\bm{E}_{\text{cry}}=R\cdot \bm{E}_{\text{lab}}$ (Fig.~\ref{Figure_1}e), where $R=Z_1(\varphi)X_2(\vartheta)Z_3(\gamma)$ is the Euler-transformation matrix for a rotation around the original $z$-axis ($Z1 = z$) by $\varphi$, the new $x$-axis ($X_2 = o$) by $\vartheta$ and final $z$-axis ($Z_3 = k$) by $\gamma$ (see Supplementaries~II). In the crystal frame (Fig.~\ref{Figure_1}d), the beams are decomposed into their components along the $o$ and $e$ axes and each polarisation combination of the pump ($oo,eo,oe,ee$) generates a second-harmonic field $E^u_{\text{gen}}(2\omega,X_n)$ (blue) along $u\in \{o,e\}$, which is given at the end of the $n^{th}$ single grain according to:
    \begin{equation}
            E^u_{\text{gen}}(2\omega, X_n)  = \sum_{v,w}\frac{i(2\omega) ^2 }{2\epsilon_0 c^2 k^u_3} P^{u,vw} \bigg(\frac{e^{i\Delta k^{u,vw}X_n}-1}{i\Delta k^{u,vw}}\bigg)e^{ik^u_3X_n}
    \end{equation}
    With the phase mismatch $\Delta k^{u,vw} = k^v(\omega) + k^w(\omega) - k^u(2\omega)$, the wave vector $k^u_3$ of the second-harmonic along $u$, and $P^{u,vw} = \langle\hat{\text{\bf{e}}}^{u}, \bm{P}^{vw}\rangle$ the second-harmonic polarization along $v,w\in\{o,e\}$ projected onto the $o/e$-axis ($\hat{\text{\bf{e}}}^{u}$ unit vector along $u\in\{o,e\}$), where $P_i^{vw} = 2\epsilon_0\sum_{jk}d_{ijk}E_j^vE_k^w$. The second-order nonlinear tensor $d_{ijk}$ is used in its contracted matrix form to calculate $P_i^{vw}$ (see Supplementaries~II).
    The pump $\bm{E}(\omega,0)$, as well as the total second-harmonic from all previous crystals $\bm{E}_{\text{pre}}(2\omega,0)$ are propagated through the grain. For both beams, the components along the $o$ and $e$ axis are propagated separately: $E^u(\omega,X_n) = E^u(\omega,0)e^{ik^u(\omega)X_n}$. At the end of the grain, the second-harmonic from the previous grains is summed with the second-harmonic generated in the current grain $E^u_{\text{tot}}(2\omega,X_n) = E^u_{\text{pre}}(2\omega,X_n) + E^u_{\text{gen}}(2\omega,X_n)$ (* in Fig.~\ref{Figure_1}d). The amplitude, polarisation and phase of the fields is explicitly considered in every step. This way, the model can account for interference effects. The fields of the propagated pump $E^u(\omega,X_n)$ and the total second-harmonic field $E^u_{\text{tot}}(2\omega,X_n)$ are transformed back into the lab frame, forming the output of the $n^{th}$ grain. To simulate the propagation of the light through the stick, we implemented a folding algorithm, which iteratively processes the sequence of grains. In each step, it takes the data of the current element (grain~$n$) of the sequence and the result of the previous step (grain $n-1$) as input. The information is processed to generate the output of the $n^{th}$ grain which then forms the input of the $(n+1)^{th}$ grain. The SHG intensities of each stick $I_{\text{stick}} = c \epsilon_0 (|E^a_{\text{tot}}|^2 + |E^b_{\text{tot}}|^2)/ 2$ are averaged to yield the intensity of the cuboid. The model assumes an undepleted pump, no absorption within the grains, no reflections at the grain-to-grain interfaces, and the propagation only in the forward direction (no walk-off angle).

    As an initial test of the algorithm, we computed the SHG from structures characterised by mono- and polydispersed grain sizes without randomisation of the grain orientation. The system behaves identical to a single crystal, periodically showing a totally constructive and destructive SHG along the length of the system. As expected, there is no dependence on the size distribution of the grains (see Supplementaries~III). When introducing disorder in grain orientation and grain size, the SHG intensity of the single stick depends strongly on the specific disorder configuration. In this case,  the SHG interference along the stick corresponds to the trajectory of a single random walk in the complex plane, as depicted in Fig.~\ref{Figure_1}c. As such, the SHG intensity of a single stick cannot be correlated with the number of grains $N$. Only upon averaging the intensities of multiple sticks (i.e. along the transverse dimensions of the cuboid), the total SHG intensity grows linearly with $N$, providing evidence of RQPM~\cite{vidal2006generation}. To quantitatively study the influence of birefringence, we introduce the \textit{isotropic analogue} of a birefringent material, defined as the material that has the same $\chi^{(2)}$-tensor as the original material, but with $\bar{n}_e = n_o$. Hence, the coherence length of the isotropic analogue \Lc\ is not angle dependent. This stands in contrast to the coherence length in a birefringent material \Lcbir, which depends on the angle, as well as the combination of beam polarizations. Thus, a random distribution of the grain orientations does not affect the coherence length of isotropic grains, while it randomises the coherence length of birefringent grains.
    %
    %
    \section{Results}
    \textit{Monodispersed Assemblies.---}
    \begin{figure}[tb!]
        \centering
        \includegraphics[scale=1]{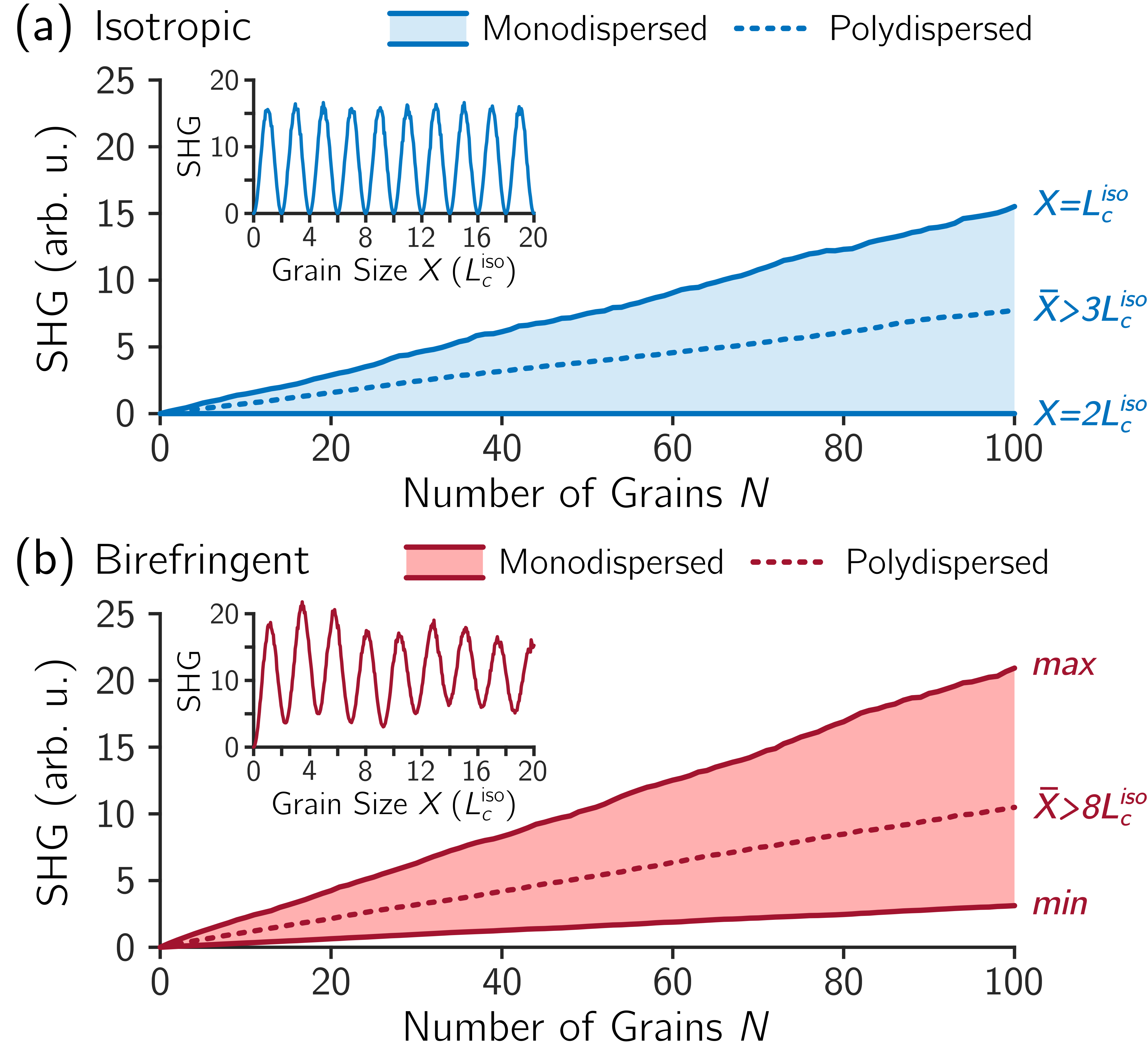}
        \caption{Scaling of the SHG intensities with the number of grains $N$ in birefringent LiNbO$_3$ and in its isotropic analogue at a pump wavelength of 930~nm. SHG intensities are averaged over 1000~sticks, leading to a linear growth. Its slope represents the SHG efficiency for a certain grain size, and the shaded area shows the possible efficiencies in monodispersed assemblies. The insets explicitly show the SHG intensity in monodispersed assemblies only as a function of the grain size~$X$ for $N=100$. In poly\-dispersed assemblies ($\sigma = 30\%$) with large grains ($\bar{X}>3$~\Lc\ in the isotropic and $\bar{X}>8$~\Lc\ in the birefringent material) the efficiency has a stable value corresponding to the dotted line. (a) The isotropic analogue has zero-efficiency minima when $\bar{X}$ is an even multiple of \Lc, and efficiency maxima when $\bar{X}$ is an odd multiple of \Lc. (b) The birefringent assembly has its minimum at $\bar{X}=9.25$~\Lc\ and its maximum at $\bar{X}=3.47$~\Lc.}
        \label{Figure_2}
    \end{figure}
    One noteworthy result of this study, which is not exclusively related to birefringence, is that even monodispersed structures (without size randomisation) sustain RQPM, as long as the grains are randomly oriented. This is shown in Fig.~\ref{Figure_2}a-b for LiNbO$_3$. The SHG efficiency per grain is measured by the slope of the linear growth, which depends on the grain size $X$. Its variation is represented by the shaded area and can be seen in the insets of Fig.~\ref{Figure_2}a-b, which report the SHG intensity as a function of the grain size. The isotropic analogue, Fig.~\ref{Figure_2}a,  shows a maximal efficiency when $X$ is an odd multiple of \Lc\ and a minimal (zero) efficiency when $X$ is an even multiple of \Lc. The latter case corresponds to the totally destructive SHG interference within the single grains. This is consistent with the results of Vidal. et al.~\cite{vidal2006generation} at low polydispersity.
    The effect of the birefringence is shown in Fig.~\ref{Figure_2}b. The maximal as well as the minimal efficiency is increased compared to the isotropic analogue. In particular, the lower bound of the cone has a non-zero slope and the SHG never vanishes. Indeed, each birefringent grain has an orientation-dependent coherence length \Lcbir, prompting that there is no specific grain size at which total destructive interference occurs for all grains. BaTiO$_3$ shows the same effect, but with a narrower cone of possible efficiencies (see Supplementaries~IV).
    The considered case of monodispersed assemblies highlights that polydispersity is neither a sufficient nor a necessary condition to achieve RQPM. This stands in contrast to the random orientation of grains, which plays a key role in reaching the RQPM regime.
    %
    
\begin{figure}[t!]
       \centering
       \includegraphics[scale=1]{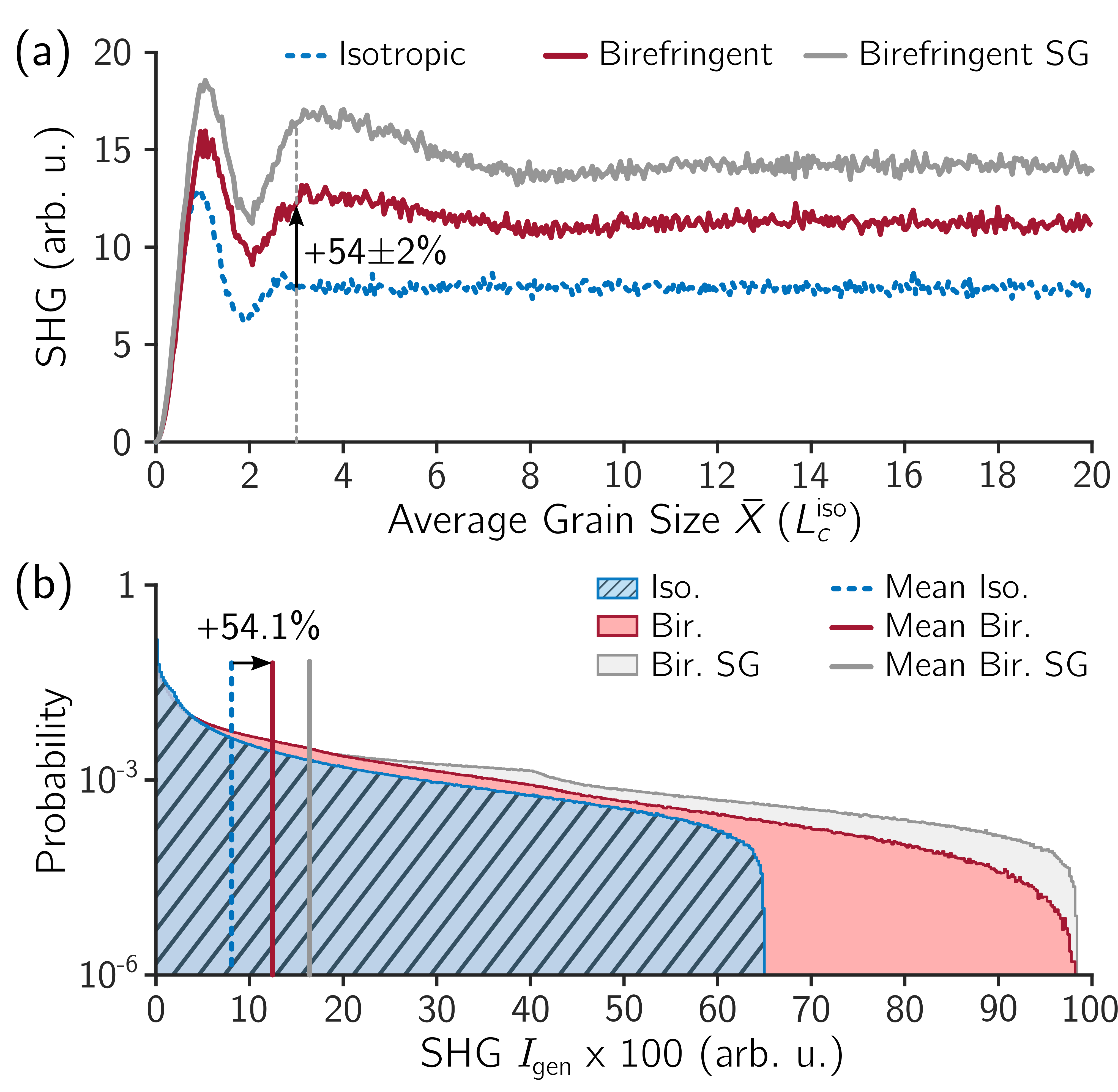}
       \caption{(a) Scaling of the SHG intensity with the average grain size for disordered assemblies of isotropic LiNbO$_3$, birefringent LiNbO$_3$, and birefringent LiNbO$_3$ in the single grain (SG) approximation (single grain average $\times N$). The assemblies have a grain size polydispersity $\sigma = 30$\%, a fixed number of grains per stick $N = 100$, an averaging over 1000 sticks, and are simulated at a pump wavelength of 930 nm. At this wavelength, phase matching in LiNbO$_3$ is not possible. 
       (b)~Probability distributions of the SHG intensities generated by the individual grains within the assembly ($\bar{X}$ = 3~\Lc \ and $\sigma = 30$\%) for isotropic LiNbO$_3$, birefringent LiNbO$_3$, and birefringent LiNbO$_3$ in the single grain (SG) approximation ($N=1$, linearly polarized input beam). Probabilities are calculated as the relative frequency of appearance of an SHG intensity value over $10^7$ grains. The SHG binning is 0.002. The SHG intensities are multiplied by 100 to be directly comparable with the intensities of the 100-grain-assemblies in (a).}
       \label{Figure_3}
    \end{figure}

    \textit{Polydispersed Assemblies.---}
    In Fig.~\ref{Figure_3}a the SHG intensity of LiNbO$_3$ assemblies (red) and of their isotropic analogue (blue) is shown for an increasing average grain size $0 \leq \bar{X} \leq 20$~\Lc , a polydispersity $\sigma = 30\%$, and a fixed number of grains $N = 100$. Both curves follow the expected trend for RQPM~\cite{vidal2006generation}. In the small grain regime ($\bar{X}>$~\Lc) the SHG intensity grows  with the grain size without an appreciable difference between the birefringent and the isotropic case. Maximal intensities are reached when $\bar{X}=$~\Lc, for the isotropic analogue and for slightly larger grains in birefringent LiNbO$_3$. After an initial peak, the SHG gets independent of the average grain size, at $\approx3$~\Lc\ for the isotropic analogue and at $\approx8$~\Lc\  for the birefringent LiNbO$_3$. In this stabilised large-grain regime, the SHG intensity of the birefringent LiNbO$_3$ is significantly larger than the stable value of the isotropic analogue. Specifically, we observe a peak enhancement of $54\pm2$\% at 3 \Lc. The effect is weaker in BaTiO$_3$, where the birefringence leads to an SHG enhancement of $9.25\pm0.8$\% (see Supplementaries~IV). The enhancement can also be seen on the slope of the of dotted lines in Fig.~\ref{Figure_2}a-b, which report the linear scaling of the SHG intensity for assemblies with polydispersed grains at large grain sizes. 
    
    The origin of this overall enhancement lies in the distribution of the SHG intensities generated by the individual grains $I_{\text{\text{gen}}} = c \epsilon_0(|E^o_{\text{gen}}|^2+|E^e_{\text{gen}}|^2)/2$ within the cuboid, which are shown in Fig.~\ref{Figure_3}b for both the birefringent (red) and the isotropic (blue) LiNbO$_3$. 
    These two SHG intensity distributions have been obtained by extracting $I_{\text{\text{gen}}}$ of the individual grains from the complete simulation depicted in Fig.~\ref{Figure_1}d, thus considering the specific polarization state of the pump (phase retardation between the $o$ and $e$ components) that the propagation through the previous birefringent grains introduced.
    The birefringence significantly widens the range of possible SHG intensities in the individual grains, due to the increased coherence lengths \Lcbir\ $ > $~\Lc \ for certain grain orientations. Consequently, the average SHG intensity of the individual grains of birefringent LiNbO$_3$ is $54.1\pm0.3$\% higher than the isotropic analogue. Similarly for BaTiO$_3$, the enhancement is $9.24\pm 0.1$\%, whose single-grain SHG intensity distribution is shown in the Supplementaries~IV. The enhancements of $I_{\text{gen}}$ are consistent with the enhancement of the assembly calculated by considering the vectorial propagation of all components.
    We highlight that the birefringence is only beneficial in grains larger than the isotropic coherence length $X_n > $~\Lc. In smaller grains, the grain size $X_n$, rather than the coherence length, is the length scale limiting their SHG intensity (see Supplementaries~V).
    To underline the importance of modeling the phase propagation between the grains, we calculated the SHG scaling also in the single grain (SG) approximation, i.e. by considering randomly oriented grains that are independently illuminated with a linearly polarized pump. Their average SHG is then multiplied by $N$ to get the SHG efficiency of the assembly $I^{\text{SG}}_{\text{tot}} = N \times \langle I^{\text{SG}}_{\text{gen}} \rangle$. As shown in Fig.~\ref{Figure_3}a (grey line), the SG approximation overestimates the SHG enhancement in the birefringent assembly of LiNbO$_3$ almost by a factor 2. This is because the SG approximation ignores the retardation introduced in the pump and hence calculates an inaccurate distribution of single grain intensities, as shown in Fig.~\ref{Figure_3}b (grey area). Isotropic materials, on the other hand, can be accurately simulated with the SG approximation, since they do not modify the polarization state of the pump.
    %
    
\begin{figure}[tb!]
        \centering
        \includegraphics[scale=1]{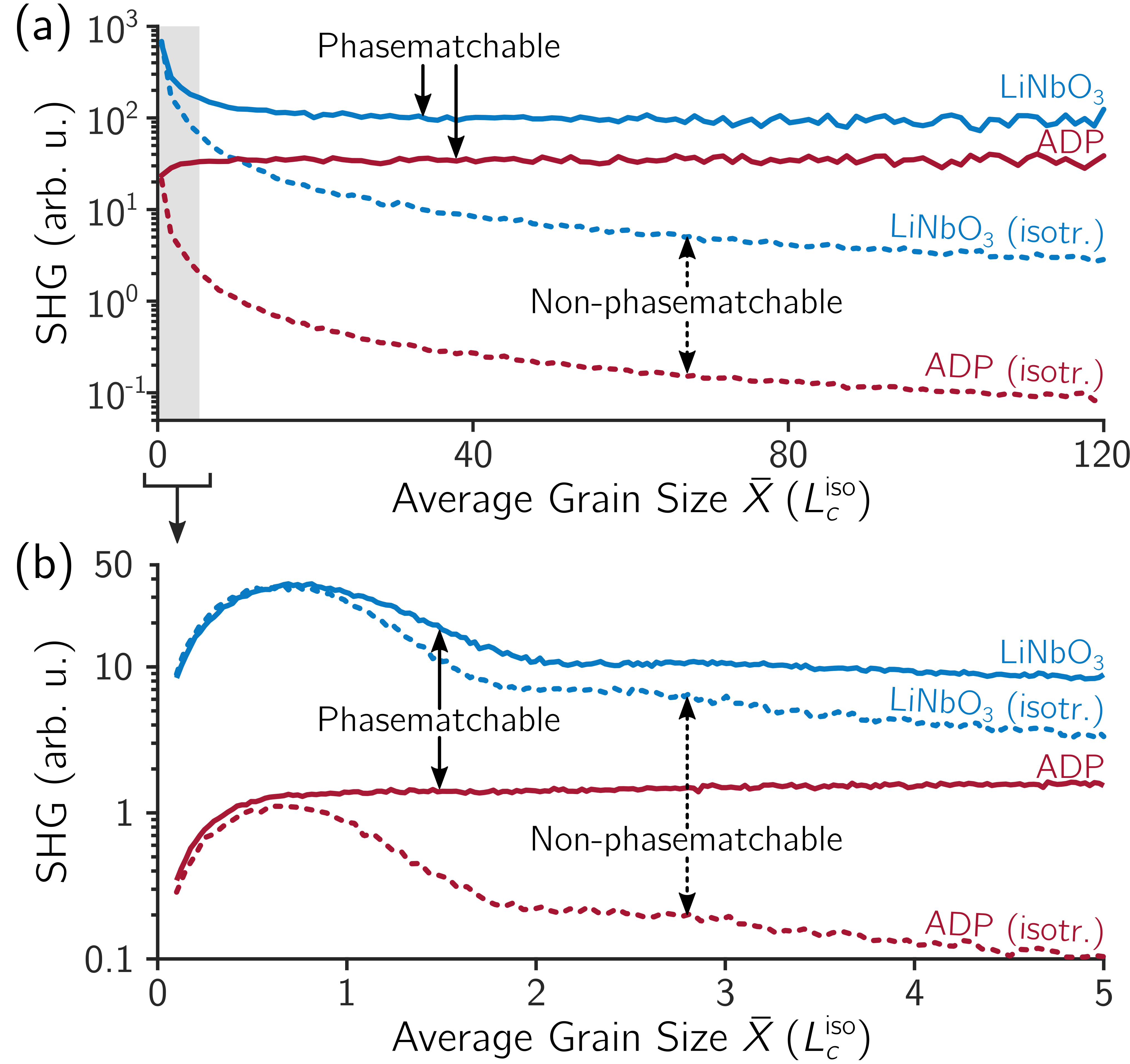}
        \caption{Scaling of the SHG intensity with the average grain size $\bar{X}$ in disordered assemblies of LiNbO$_3$ and ADP and their isotropic analogues at a pump wavelength of $\lambda = 1200$~nm. At this wavelength, phase matching is possible in both materials. The total length of the sticks is kept constant and the grains have a polydispersity $\sigma = 30$\%.
        (a) The total length of the sticks is fixed at 1000~\Lc. When changing $X$, the number of grains per stick varies between 1000 grains (at $\bar{X} = 1$ \Lc) to  10 grains (at $\bar{X} = 100$ \Lc). The phasematchable materials show a different scaling in comparison to their not phasematchable isotropic analogues. Non-phase-matchable LiNbO$_3$ can also be simulated by using a shorter pump wavelength which is shown in the Supplementaries~VI. 
        (b) Higher resolution simulation of the small grain size regime, performed at a reduced stick length of 50~\Lc\ for computational efficiency. The shift in the y-scale compared to (a) is due to the different stick length. In this small grain regime the scaling of the SHG intensity is not determined solely by the phase-matchability, but becomes material specific.}
        \label{Figure_4}
    \end{figure}

    \textit{Polydispersed Assemblies under Phase-matchable Con\-di\-tions.---}
    So far, we have performed our analysis in a non-phase-matchable regime. This means that the wavelengths were chosen such that there is no crystal orientation for which phase matching is possible in the studied materials. Since our model considers birefringence, we can numerically investigate the phase-matchable regime in RQPM. For this, we consider LiNbO$_3$ at 1200~nm (see Supplementaries~I) and compare it to the non-phase-matchable regime by using the corresponding isotropic analogue.
    We investigate how the SHG scaling with the average grain size $\bar{X}$ is affected by the phase-matchability of the grains, since this scaling behavior is considered experimentally  to discriminate between phase-matchable and non-phase-matchable materials by using powder samples. This method was first introduced in the seminal work of Kurtz and Perry (KP)~\cite{Kurtz1968/doi:10.1063/1.1656857} and recently revised by Aramburu et. al.~\cite{Aramburu2013Second}. 
    The calculated SHG scalings with increasing average grain size $\bar{X}$ for LiNbO$_3$ and for its isotropic analogue are shown in Fig.~\ref{Figure_4}a. To be comparable with KP, the length of the cuboid is kept fixed, which reduces the total number of grains when increasing the grain size. This re-scales the SHG efficiency of Fig.~\ref{Figure_3} by 1/$\bar{X}$, e.g. the constant SHG efficiency of the isotropic analogue from Fig. \ref{Figure_3} scales with 1/$\bar{X}$ in Fig.~\ref{Figure_4}. Another effect of the constraint on the total length of the cuboid, is the shift in the most efficient grain size from $\bar{X} \sim 1$ \Lc\ to $\bar{X} \sim 0.7$ \Lc.
    For comparison with the results reported by KP~\cite{Kurtz1968/doi:10.1063/1.1656857}, we performed the same calculations on (Ammonium Dihydrogen Phosphate) ADP, which are also shown in Fig.~\ref{Figure_4}a. Our model, is able to reproduce the results obtained experimentally by KP in the large-grain regime ($\bar{X}>5$~\Lc). Namely, in the birefringent and phase-matchable case, the SHG intensity is independent of $\bar{X}$, while we observe a $1/\bar{X}$ dependence for its non-phase-matchable isotropic analogue for both ADP and LiNbO$_3$. This implies that the enhancement of the SHG intensity, which is introduced by the birefringence, grows linearly with the average grain size as soon as some of the grains can be phase matched. This linear dependence is clearly visible when considering the SHG scaling without constraint on the cuboid length and with a fixed number of grains (Supplementaries~VI).
    
    Notably, the scaling behavior of the two materials is very different in the small-grain regime ($\bar{X}<5$~\Lc), as shown in Fig.~\ref{Figure_4}b. For ADP, one can see the monotonic increase in the phase-matchable case, while the non-phase-matchable case shows an SHG peak followed by a decreasing trend, in agreement with the results of KP~\cite{Kurtz1968/doi:10.1063/1.1656857}. Surprisingly, for LiNbO$_3$, there is no significant difference between the phase-matchable and the non-phase-matchable case for $\bar{X}<5$~\Lc. Both show an initial peak followed by a decreasing trend and they could not be distinguished by a scaling experiment. This result shows that for average grain sizes $\bar{X}$ smaller than 5~\Lc\ the SHG scaling discovered by KP is not universal, but rather becomes material specific. 
    In LiNbO$_3$, the source of the deviation from the phase-matchable scaling of KP can be found in the largest tensor element $d_{33}$ which determines the $E_z\cdot E_z$ component of the SHG. This is a non-phase-matchable tensor element since the electric field of the SHG is orthogonal to the $\bm{z}$-axis for the ordinary beam component, and hence its contribution is zero in the phase-matchable configuration ($oo$-$e$). 
    However, for randomly oriented grains smaller than \Lc, the contributions of the $d_{33}$ tensor element dominate the total SHG. For such small grains the maximum SHG intensity is limited by the grain size rather than the coherence length. At larger grain sizes, the smaller---but phase-matchable---coefficient starts to contribute significantly to the total SHG and leads to the KP behaviour. In ADP, all tensor components are the same, and therefore influence the phase-matchable and the non-phase-matchable SHG in equal parts. We conclude that only when the phase-matchable tensor components are much smaller than the non-phase-matchable ones, the disordered assembly behaves like the non-phase-matchable case of KP in the small grain regime.
    %
    %
    
    \section{Conclusions}
    We have developed a vectorial model that is capable of calculating the second-harmonic generation through RQPM in any transparent, birefringent, $\chi^{(2)}$-disordered medium. This gives the presented model a wide range of applicability: from single grains, to disordered assemblies, to assemblies with specific size and orientation correlations between the grains. We employ the presented model to explore the effects of the birefringence on the RQPM process, by considering fully disordered assemblies of LiNbO$_3$, BaTiO$_3$, and ADP in both non-phasematchable and phase-matchable conditions. We pinpoint the role played by the birefringence through a comparison with isotropic analogue materials having $\bar{n}_e = n_o$. While no influence of the birefringence is appreciable for smaller grains, the birefringence starts to play a notable role once the grains in the system are larger than the isotropic coherence length \Lc\ in three studied situations (monodispersed, polydispersed, and polydispersed with phase matching). 
    We identify the random orientation of the grains as the key feature to enter the RQPM regime, a property enabling RQPM even in monodispersed polycrystalline assemblies. With a proper choice of grain size, the monodispersed assembly can outperform the SHG efficiency of a polydispersed assembly in the large grain regime. This property reveals the potential of the nonlinear generation from layered structures (similar to periodically poled ones) which would require to control only the layer thickness, without restrictions in the crystal orientation. In this monodispersed case, the birefringence introduces a randomisation of the coherence length that relaxes the grain size dependence of the SHG intensity. In polydispersed assemblies, the birefringence of the material leads to a grain size independent efficiency increase of up to 54$\%$ in comparison to the corresponding isotropic analogue. This efficiency increase is ascribed to an increase in the average grain efficiency, stemming from larger coherence lengths for certain grain orientations.
    We show that the explicit grain-to-grain propagation of the polarization components is necessary for the accurate prediction of this SHG efficiency enhancement and in general for the accurate description of the three-wave mixing process in disordered birefringent media.
    We show that the Kurtz and Perry method to discriminate between phase-matchable and non-phase-matchable materials, cannot be applied when the grains are smaller than the coherence length, e.g. nano-powders. In this small grain regime, the dependence of the SHG on the average grain size is given by the specific $\chi^{(2)}$ tensor of the grains and thus becomes material specific.
    %
    %
    Our findings show how a larger set of materials, including birefringent crystals, can serve for RQPM applications and clarify the limitations of a widely-applied method for the characterization of nonlinear optical materials. Due to the non resonant nature of the SHG and the RQPM, we expect our results to be valid in a wide range of wavelengths (e.g. hundreds of nanometers for metal-oxides), as long as the absorption is negligible.
    The model could be extended to include absorption within the grains, scattering effects at the interfaces, and electro-optic effects. Moreover, it could be generalized to study other three-wave mixing processes such as sum- and difference-frequency generation.

\section{Acknowledgements}
    This research has received funding from the European Union’s Horizon 2020 research and innovation program under the Marie Skłodowska-Curie grant agreement no. 800487 (SECOONDO) and from the European Research Council under the grant agreement no. 714837 (Chi2-nano-oxides). We thank the Swiss National Science Foundation (SNF) grant 150609.

\bibliography{main}

\end{document}


\preprint{APS/123-QED}

    \title{SUPPLEMENTARY INFORMATION \texorpdfstring{\\}{} Modeling of Random Quasi-Phase-Matching in Birefringent Disordered Media}
    
\author{Jolanda S. Müller}
\author{Andrea Morandi}
\author{Rachel Grange}
\author{Romolo Savo}
\email{savor@phys.ethz.ch}

\affiliation{Optical Nanomaterial Group, Institute for Quantum Electronics, Department of Physics, ETH Zurich, Zurich, Switzerland}

\date{\today}

    \maketitle
    \tableofcontents

\newpage
\section{Material Parameters}
    \label{sec:materialcoefficients} 
    \subsection{Nonlinear Optical Tensors and Refractive Indices}
        \subsubsection*{Lithium Niobate}
            LiNbO$_3$ is trigonal, point group $3m$ ($C_{3v}$). The reduced $\chi^{(2)}$ tensor of LiNbO$_3$ is given in Eq. (\ref{LNO_tensor})~\cite{LNO_tensor}. The ordinary and extraordinary refractive indices are calculated at room temperature (T~=~20~\textdegree C) according to the Sellmeier equation~\cite{LNO_ref}. At a wavelength of 930 nm the refractive indices are $n_o = 2.2436$ and $\bar{n}_e = 2.1634$.
            
            \begin{equation}
                \label{LNO_tensor}
                d_{\text{LNO}} =
                    \begin{bmatrix} 
                        0 & 0 & 0 & 0 & d_{31} & -d_{22} \\
                        -d_{22} & d_{22} & 0 & d_{31} & 0 & 0 \\
                        d_{31} & d_{31} & d_{33} & 0 & 0 & 0 \\ 
                    \end{bmatrix}
                \quad\textrm{with}\quad
                    \begin{matrix}
                        d_{22} = &+ 2.1 & \text{ pm/V}\\
                        d_{31} = &- 4.2 & \text{ pm/V}\\
                        d_{33} = &-27.0 & \text{ pm/V}\\
                    \end{matrix}
            \end{equation}
        
        \subsubsection*{Barium Titanate}
            BaTiO$_3$ is tetragonal, point group $4mm$ ($C_{4v}$).
            The reduced $\chi^{(2)}$ tensor of BaTiO$_3$ is given in Eq. (\ref{BTO_tensor})~\cite{BTO_tensor}. The ordinary and extraordinary refractive indices at different wavelengths are calculated according to the Sellmeier equation~\cite{BTO_ref}. At a wavelength of 930~nm the refractive indices are $n_o = 2.3502$ and $\bar{n}_e = 2.3079$.
            \begin{equation}
                \label{BTO_tensor}
                d_{\text{BTO}} =
                    \begin{bmatrix} 
                        0 & 0 & 0 & 0 & d_{15} & 0 \\
                        0 & 0 & 0 & d_{15} & 0 & 0 \\
                        d_{31} & d_{31} & d_{33} & 0 & 0 & 0 \\ 
                    \end{bmatrix}
                \quad\textrm{with}\quad
                \begin{matrix}
                    d_{15} = &-17.0 & \text{ pm/V}\\
                    d_{31} = &-15.7 & \text{ pm/V}\\
                    d_{33} = &- 6.8 & \text{ pm/V}\\
                \end{matrix}
            \end{equation}
        
        \subsubsection*{Adenosine Diphosphate}
            ADP is tetragonal, point group $\overline{4}2m$ ($D_{2d}$). The reduced $\chi^{(2)}$ tensor of ADP is given in Eq. (\ref{ADP_tensor})~\cite{martienssen2006springer}. The ordinary and extraordinary refractive indices at different wavelengths are calculated according to the Sellmeier equation~\cite{Zernike:64}. For example at a wavelength of 930 nm the refractive indices are $n_o = 1.5114$ and $\bar{n}_e = 1.4708$.
            
            \begin{equation}
                \label{ADP_tensor}
                d_{\text{ADP}} =
                    \begin{bmatrix} 
                        0 & 0 & 0 & d_{36} & 0 & 0 \\
                        0 & 0 & 0 & 0 & d_{36} & 0 \\
                        0 & 0 & 0 & 0 & 0 & d_{36} \\ 
                    \end{bmatrix}
                \quad\textrm{with}\quad
                    d_{36} =  0.47 \text{ pm/V}
            \end{equation}
    

    \subsection{Phase Mismatch and Coherence Length}
        \label{sec:simulation:theory:phasemismatch}
\begin{figure}[ht!]
            	\centering
            	\includegraphics{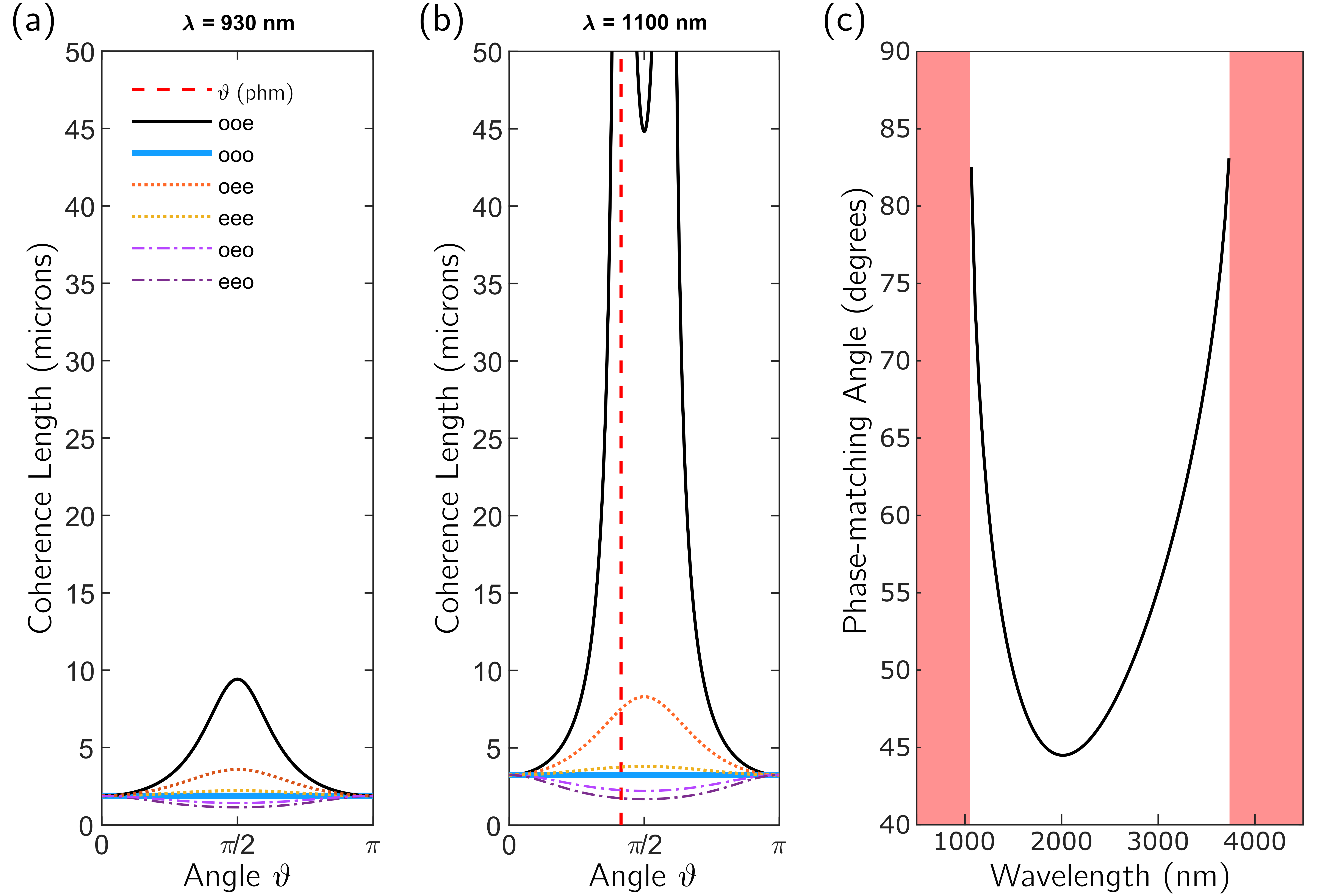}
            	\caption{(a) Coherence lengths in LiNBO$_3$ for different angles $\vartheta$ at 930~nm. At this wavelength all coherence lengths are finite and phase matching is not possible. (b) Coherence lengths in LiNBO$_3$ for different angles $\vartheta$ at 1100~nm. When the phase-matching condition is fulfilled, the coherence length gets infinitely large, as indicated with the red dashed line at $\theta$ (phm). (c) The phase-matching angle for different wavelengths in LiNbO$_3$. The red shaded area indicates the wavelength range in which the phase matching condition does not have a solution. }
            	\label{img:coherencelength}
        \end{figure}

        For isotropic materials, the phase mismatch between the funamental beam ($\omega$) and the SHG ($2\omega$) is independent of the beam polarizations and the orientation of the crystal and it is given as
        \begin{equation}
            \Delta k = 2 |\bm{k}(\omega)| - |\bm{k}(2\omega)| =  \frac{2\omega}{c}(n_{\omega} - n_{2\omega})  
        \end{equation}
        with the wave vector $\bar{k}$ and the refractive indices $n_{\omega}$ and $n_{2\omega}$ of the fundamental beam and the SHG, respectively. The coherence length is the length at which the relative phase shift of the generated waves reaches $\frac{\pi}{2}$, and is defined as \Lc $= \pi / \Delta k$.
        In a birefringent crystal the refractive indices depend on the the beam polarization, and on the angle between the optic axis of the crystal and the wave vector $\bm{k}$. In this case, the general expression of the phase mismatch between the fundamental and the second harmonic $\Delta k$ takes into account the different polarisation combinations. This leads to $\Delta k^{u,vw} = k^v(\omega) + k^w(\omega) - k^u(2\omega)$ with $u,v,w \in \{o,e\}$ and is written as eight different values for $\Delta k$ depending on the beam combinations:
        $\Delta k^{o,oo}$, $\Delta k^{o,ee}$, $\Delta k^{o,oe}$, $\Delta k^{o,eo}$, $\Delta k^{e,oo}$, $\Delta k^{e,ee}$, $\Delta k^{e,oe}$, $\Delta k^{e,eo}$.
        Consequently, the coherence length for each beam polarization combination is derived from the corresponding phase mismatch $L_c^{\text{bir}}(\vartheta) = \pi / \Delta k^{u,vw}$ and it depends on $\vartheta$ as depicted in Fig.~\ref{img:coherencelength}. We note that beam combinations of type $e,oo$ (solid black) have the largest coherence length, and the beam combination $o,oo$ (solid blue) has an angle independent coherence length. At 930~nm, LiNbO$_3$ has a coherence length in the interval [1.14~$\mu$m, 9.40~$\mu$m], with the constant $o,oo$-value of 1.88~$\mu$m and an average value of 2.314~$\mu$m. The coherence length of BaTiO$_3$ lies within the interval [0.95~$\mu$m,1.87~$\mu$m], with a constant $o,oo$-value of 1.57~$\mu$m and an average value of 1.23~$\mu$m.
        
    The coherence length becomes infinitely large $L_c \rightarrow \infty$, when the phase matching condition $\Delta k = 0$ is satisfied, which means that both waves travel through the medium at the same speed. To allow for phase matching the refractive index of the fundamental wave ($n_{\omega}$) must be equal to the refractive index of the SHG ($n_{2\omega}$). Considering the angle dependent refractive indices of both frequencies the phase-matching condition becomes~\cite{boyd2008nonlinear}:
    
    \begin{equation}
    \label{eq:phasematching-angle}
        \frac{\sin^2\vartheta}{\bar{n}_e(2\omega)^2} + \frac{\cos^2\vartheta}{n_o(2\omega)^2}  =  \frac{1}{n_o(\omega)^2}
    \end{equation}
    
    While the solution of this equation provides the wavelength dependent phasematching angle, it is not guaranteed that a solution for every wavelength exists. For LiNbO$_3$ a solution only exists between 1065 nm to 3732 nm. The corresponding phase-matching angles are reported in Fig.~\ref{img:coherencelength}b, which shows the phase-matching angles from 500 nm to 4500 nm. It is possible that the birefringence of a material is not sufficient to allow for the solution of Eq.~\ref{eq:phasematching-angle} at any wavelength. This is the case of BaTiO$_3$ which is therefore never phase matchable.

\newpage
\section{Algorithm Details}
\begin{figure}[ht!]
            \centering
            \begin{minipage}{0.6\textwidth}
                 \begin{verbbox}[\mbox{}]
                 
    Grains = [G1, G2, ..., GN]
    Beams = InputBeams
     
    for i = 1 to N
        Beams = f(Beams, Grains(i))
        TotalBeams(N) = Beams
    end
          
    OutputBeams = Beams
                \end{verbbox}
                \par\colorbox{codegray}{\parbox{0.9\textwidth}{\theverbbox}}\par
            \end{minipage}%
            \begin{minipage}{0.4\textwidth}
                \centering
                \includegraphics[width=1\linewidth]{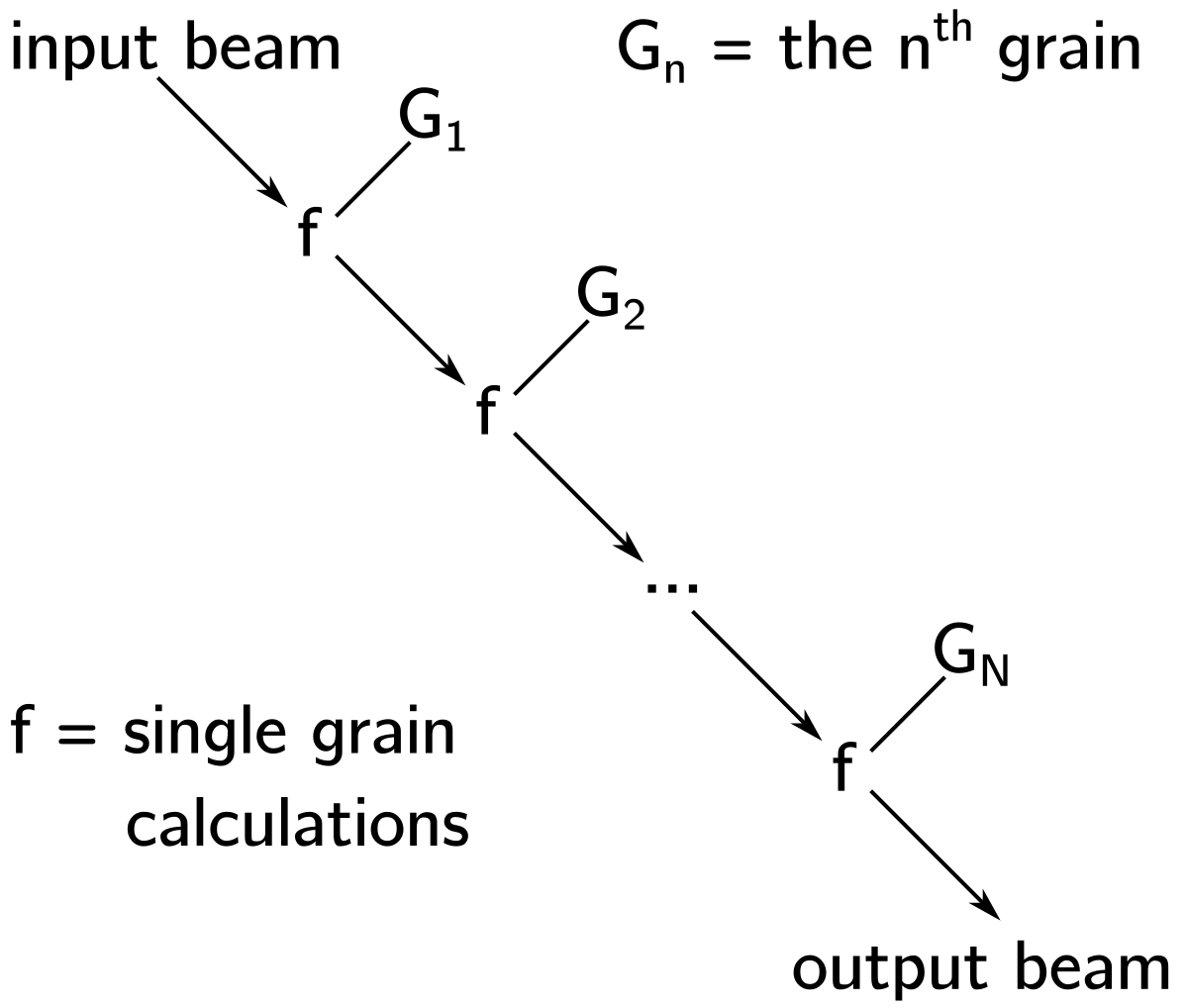}
            \end{minipage}
            \mbox{} \\
            \textbf{Single Grain Function f:}
            \begin{minipage}{1.0\textwidth}
\begin{verbbox}[\mbox{}]

    functions [BeamsLabOut] = f(BeamsLabIn, Grain)
        BeamCrystalIn = transform2Crystal(BeamsLabIn, Grain)
        
        BeamsCrystalOut.Fund = propagagte(BeamsCrystalIn.Fund, Grain)
        
        Eprop                = propagagte(BeamsCrystalIn.SHG, Grain)
        Egen                 = generateSHG(BeamsCrystalIn.Fund, Grain)
        BeamsCrystalOut.SHG  = Eprop + Egen
        
        BeamsLabOut = transform2Lab(BeamsOutCrystal, Grain)
    end
    
\end{verbbox}
\par\colorbox{codegray}{\parbox{1.0\textwidth}{\theverbbox}}\par
            \end{minipage}
            \caption{Pseudocode and schematic of the main functions in the algorithm.}
            \label{fig:pseudocode}
    \end{figure}

    The propagation through the stick of $N$ grains is performed with a fold function, which takes an input value (input beam), and a data structure (array of $N$ grains) and reduces them to a single output value (output beam) through recursive processing with a function $f$ (performing the single grain operations), by using the output of step $n$ as input of step $(n+1)$: $\vec{E}_{\text{in}}^{n+1}(\omega) = \vec{E}_{\text{out}}^{n}(\omega)$ for both $\omega$ and $2\omega$. The output of each grain is composed of the propagated fundamental field, and the total SHG. The propagation is computed separately for the ordinary and extraordinary components $u \in \{o,e\}$ and for both frequencies $\omega$ and $2\omega$: $E^u_{\text{prop}}(\omega,X_n) = E^u(\omega,0) \cdot e^{iX_nk^u(\omega)}$. The total SHG is given by the sum of the propagated SHG input form the previous grain and the SHG generated within the current grain: $E^u_{\text{total}}(2\omega) = E^u_{\text{gen}}(2\omega,X_n) + E^u_{\text{prop}}(2\omega,X_n)$. The computational procedure is shown in simplified pseudo-code in Fig.~\ref{fig:pseudocode}.
        
    \subsection{Reference Frame Transformations}
        The transformation matrix $R$ ($\text{c} \equiv \cos$ and $\text{s} \equiv \sin$) is given by:
        \begin{equation}
        	\label{eq:simu:trafo-matrix}
        	R =
        	\begin{pmatrix}
        		\text{c}(\varphi)\text{c}(\gamma)-\text{s}(\varphi)\text{c}(\vartheta)\text{s}(\gamma) & -\text{c}(\varphi)\text{s}(\gamma)-\text{s}(\varphi)\text{c}(\vartheta)\text{c}(\gamma) & \text{s}(\vartheta)\text{s}(\varphi) \\
        		\text{s}(\varphi)\text{c}(\gamma)+\text{c}(\varphi)\text{c}(\vartheta)\text{s}(\gamma) & -\text{s}(\varphi)\text{s}(\gamma)+\text{c}(\varphi)\text{c}(\vartheta)\text{c}(\gamma) & -\text{s}(\vartheta)\text{c}(\varphi) \\
        		\text{s}(\vartheta)\text{s}(\gamma) & \text{s}(\vartheta)\text{c}(\gamma) & \text{c}(\vartheta)\\
        	\end{pmatrix}
        \end{equation}
        
        After the transformation into the crystal frame, the E-fields are then projected onto the ordinary and extraordinary axis ($o,e$) for separate propagation. The ordinary axis is orthogonal to the plane defined by the propagation vector $\bm{k}_{\text{cry}}$ and the optic axis $\bm{z}$ and its unit vector is given by $\hat{\text{\bf{e}}}^{o} = \frac{\bm{z}\times\bm{k}_{\text{cry}}} {|\bm{z} \times \bm{k}_{\text{cry}}|}$. The extraordinary axis is orthogonal to $\bm{k}_{\text{cry}}$ and to the optic axis $\bm{z}$ with the unit vector $\hat{\text{\bf{e}}}^{e} = \frac{\bm{k}_{\text{cry}}\times\hat{\text{\bf{e}}}^{o}} {|\bm{k}_{\text{cry}}\times\hat{\text{\bf{e}}}^{o}|}$.
 
    \subsection{Non-linear Polarisation for Different Polarisation Directions}
        \label{sec:simulation:theory:polarisation}
        The non linear polarisation generated by a discrete set of incoming electric fields $\vec{E}(\omega_n)$ is given by Boyd~\cite{boyd2008nonlinear}:
        \begin{equation}
        \label{eq:polarisation1}
        P_i(\omega_n+\omega_m) = 2\epsilon_0\sum_{jk}\sum_{nm}d_{ijk}E_j(\omega_n)E_k(\omega_m)
        \end{equation}
        where $d_{ijk}$ is the material specific $\chi^{(2)}$ tensor. For the second harmonic generation there is just one fundamental frequency and therefore $\omega_n = \omega_m = \omega$. Thus the sum over $m$ and $n$ is ignored. The E-fields $E_j$ are decomposed in their ordinary and extraordinary contribution $E_j(\omega) = E_j^o(\omega) + E_j^e(\omega)$. For simplicity ($\omega$ is not written below). Thus we can consider the contributions to the polarisation by the four combinations of the ordinary and extraordinary complex fundamental $\vec{E}$-fields (including the phase):
        
        \begin{equation}
        \label{eq:polarisation2}
        \begin{aligned}
        	P_i(2\omega) &=  2\epsilon_0\sum_{jk}d_{ijk}(E_j^oE_k^o+E_j^eE_k^e+E_j^oE_k^e+E_j^eE_k^o)\\
        	 &=  P_i^{oo}+P_i^{ee}+P_i^{oe}+P_i^{eo}
        	\quad \text{where} \quad
        	P_i^{vw} = 2\epsilon_0\sum_{jk}d_{ijk}E_j^vE_k^w
        \end{aligned}
        \end{equation}
        
        Due to the Kleinmann Symmetry the tensor $d_{ijk}$ is symmetric under exchange of $j$ and $k$ and is written in its contracted matrix form $d_{il}$ ~\cite{boyd2008nonlinear} for the material coefficients, which we use with the matrix equation:

        \begin{equation}
            \begin{aligned}
                \label{project_oexyz}
                \vec{P}^{vw} =
                \begin{bmatrix} 
                P_x^{vw}(2\omega) \\
                P_y^{vw}(2\omega) \\
                P_z^{vw}(2\omega) \\
                \end{bmatrix}
                =
                2 \epsilon_0
                \begin{bmatrix}
                d_{11}&d_{12}&d_{13}&d_{14}&d_{15}&d_{16}\\
                d_{21}&d_{22}&d_{23}&d_{24}&d_{25}&d_{26}\\
                d_{31}&d_{32}&d_{33}&d_{34}&d_{35}&d_{36}\\
                \end{bmatrix}
                \begin{bmatrix}
                E_x^vE_x^w \\
                E_y^vE_y^w\\
                E_z^vE_z^w \\
                E_y^vE_z^w + E_z^vE_y^w\\
                E_x^vE_z^w + E_z^vE_x^w\\
                E_x^vE_y^w + E_y^vE_x^w\\
                \end{bmatrix}
            \end{aligned}
        \end{equation}
        
        The resulting polarisations are generally not polarised along just one axis ($o/e$). Therefore, each $\vec{P}^{vw}$ is decomposed along the $o/e$-direction within the current crystal by taking the inner product~\footnote{The inner product is only symmetric with respect to the complex conjugate $<v,w> = \overline{<w,v>}$ and since $\vec{P}^{uv}$ is a complex quantity containing phase information, the order of terms is relevant, to avoid flipping the phase by $\pi$.} with the unit vector pointing in the (extra-)ordinary direction: $P^{u,vw} = <\hat{\text{e}^{u}}, \vec{P}^{vw}>$. From this $E^u_{\text{gen}}$ is calculated according to Eq.~(1) in the main text, of which we show the derivation in the next section.

\subsection{Second Harmonic E-field at the End of One Grain (Derivation of Eq. (1))}

We start from the coupled-amplitude equation provided by Boyd~\cite{boyd2008nonlinear}, which describes how the amplitude of the non-linear wave ($\omega_3$) depends on the amplitudes of the fundamental waves ($\omega_1$ and $\omega_2$). In our case, $\omega_1 = \omega_2 = \omega$ and $\omega_3 = 2\omega$.

\begin{equation}
    \label{eq:coupled-amplitude-equation}
    \frac{dA_3}{dz} = \frac{i\omega_3}{2\epsilon_0n_3c}p_3e^{i\Delta kz}    
\end{equation}

 $A_3$ is the amplitude of the non-linear wave (envelope function). The non-linear polarisation is given by $\vec{P}_3(z,t)= p_3e^{i(k_1+k_2)z} + c.c.$. We note that any starting phases of the fundamental $E$-fields ($\varphi_1$ and $\varphi_2$) given at $z=0$, $p=0$ must be contained in $p_3$, making it a complex quantity. Analogous to Boyd's calculations~\cite{boyd2008nonlinear}, we derive the amplitude of the sum-frequency field $A_3$ at the end of a grain of length $X_n$, by integrating Eq. \ref{eq:coupled-amplitude-equation}  from $z = 0$ to $z = X_n$ and substituting $n_3 = c|k_3|/\omega_3$: 

\begin{equation}
    A_3(X_n) = \frac{i\omega_3}{2\epsilon_0n_3c} p_3 \int_0^{X_n} e^{i\Delta kz}dz = \frac{i\omega_3}{2\epsilon_0n_3c} p_3 \left( \frac{e^{i\Delta kX_n} - 1}{i\Delta k}\right)   
            = \frac{i\omega_3^2}{2\epsilon_0c^2k_3} p_3 \left( \frac{e^{i\Delta k X_n} - 1}{i\Delta k}\right)
\end{equation}

In the birefringent case, the phase-mismatch $\Delta k$ becomes $\Delta k^{u,vw}$, dependent on the polarisation directions of the fundamental beam components $v,w\in \{o,e\}$, as well as the polarisation direction of the generated wave $u\in \{o,e\}$ as established in section \ref{sec:simulation:theory:phasemismatch}. Similarly, the non-linear polarisation $p_3$ is given by the complex quantity $\vec{P}^{u,vw}$ with $u,v,w\in \{o,e\}$ as derived in the previous section. It contains the starting phase $\varphi_1$ and $\varphi_2$ of the to fundamental beam components at the beginning of the crystal. The amplitude of the non-linear field for one set of polarisation combinations is given by:

\begin{equation}
    A_3^{u,vw}(X_n) = \frac{i\omega_3^2}{2\epsilon_0c^2k_3^u} \cdot P^{u,vw}\cdot \left( \frac{e^{i\Delta k^{u,vw}  X_n} - 1}{i\Delta k^{u,vw} }\right)
\end{equation}

We retrieve the non-linear complex E-field at the end of the crystal along one polarisation direction through multiplication with the propagation term $e^{ik^uX_n}$ and summation over the different input polarisation combinations. Lastly, we substitute $\omega_3$ with $2\omega$, yielding eq. (1) presented in the main text:

\begin{equation}
    E^u_{\text{gen}}(2\omega, X_n) = \sum_{v,w}\frac{i(2\omega)^2}{2\epsilon_0c^2k^u_3}\cdot  P^{u,vw} \cdot \left( \frac{e^{i\Delta k^{u,vw}  X_n} - 1}{i\Delta k^{u,vw} }\right)\cdot e^{ik^u_3X_n}
\end{equation}

\subsection{Ensemble Averaging}

\begin{figure}[ht!]
    \centering
    \includegraphics[width=0.5\textwidth]{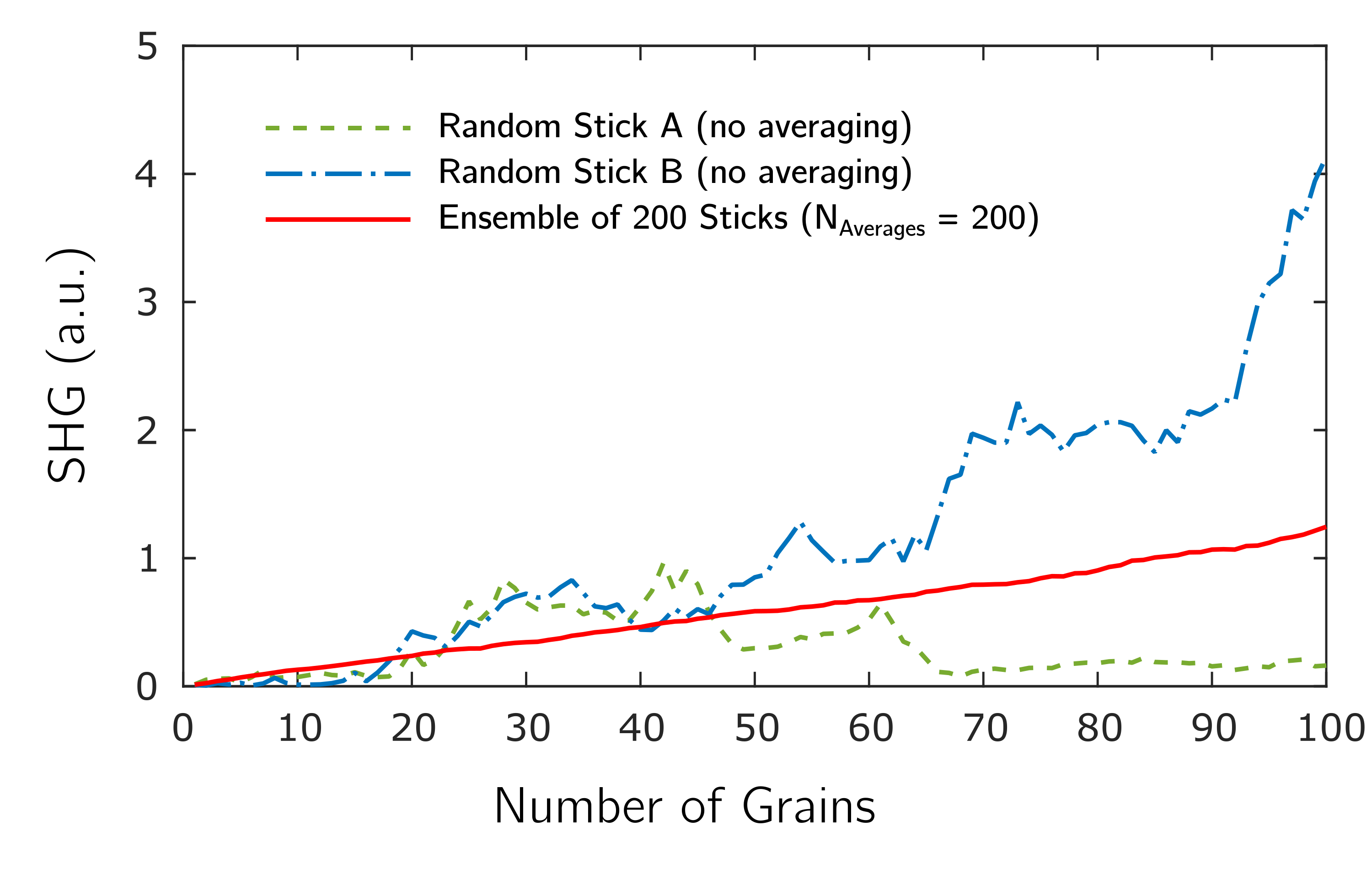}
    \caption{SHG intensity for two single sticks with independent disorder configurations (green dotted, blue dashed) and for the ensemble average over 200 sticks. After the ensemble average the SHG intensity depends linearly on the number of grains}
    \label{fig:averaging}
\end{figure}

In Fig.~\ref{fig:averaging}, we demonstrate the difference between the simulation for one single stick, and the ensemble average over multiple sticks discussed in the main manuscript. The SHG scaling with the number of grains for two sticks with independent configurations of disorder, does not show a correlation with the number of grains. The SHG of the two sticks corresponds to two different, independent random walks in the SHG complex plane. Upon taking the ensemble average over 200 sticks, the SHG grows linearly with the number of grains. Taking the ensemble average is physically justified by the fact, that the laser beam illuminates multiple parallel sticks at onces~\cite{vidal2006generation}.

\newpage
\section{Algorithm Verification}

\subsection{Visualisation of the Wave Propagation and Summation}
\begin{figure}[ht!]
    	\centering
    	\includegraphics[width=\textwidth]{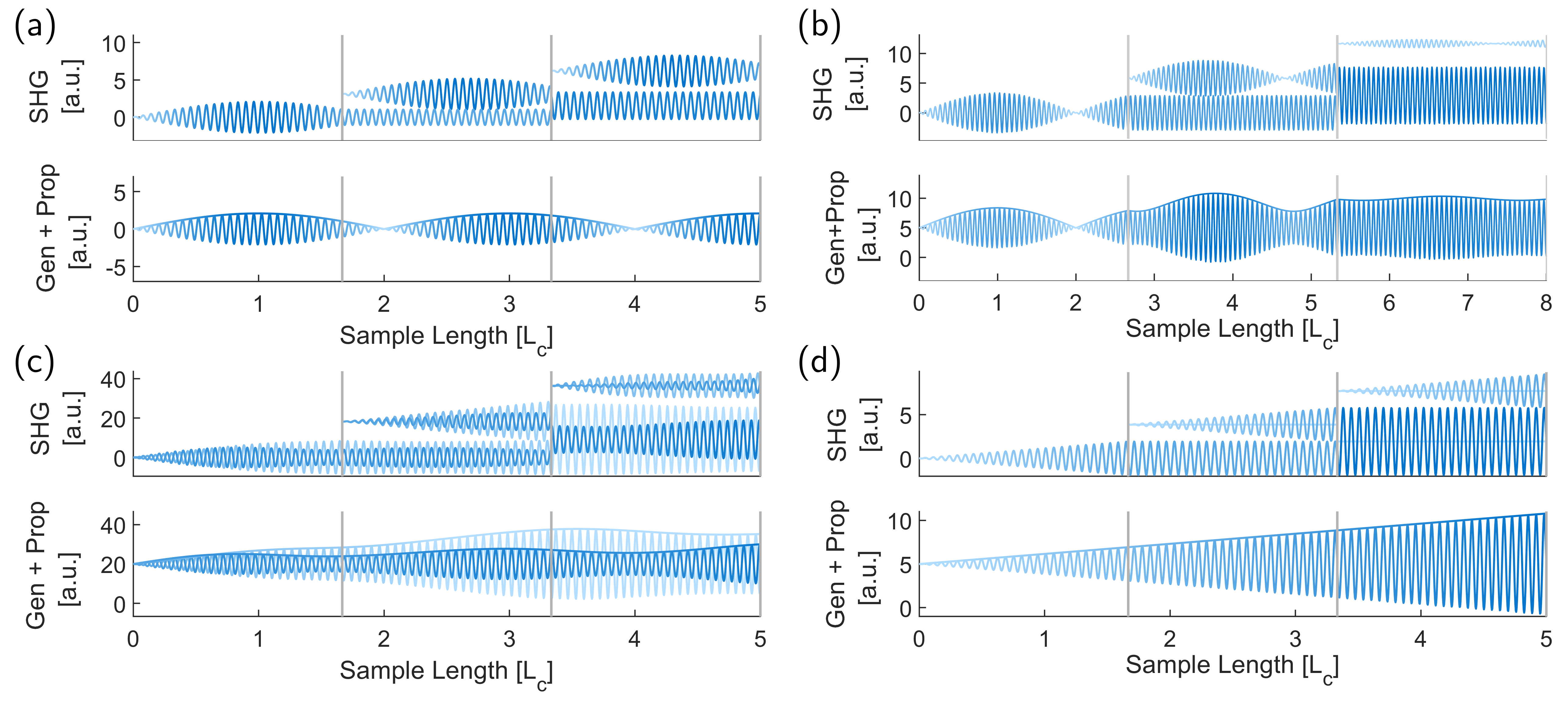}
    	\caption{For each system, the first three grains are shown, separated by a grey line. The generated and propagated waves are shown individually in the top half and their resulting wave in the bottom half of \textbf{a-d}. (a) Isotropic system with fixed orientation. The coherence length is clearly visible in the resulting wave. (b) Birefringent system with fixed orientation. The contributions along a and b in the lab frame are shown in bark/light blue. (d) Birefringent system with fixed, phasematched orientation. The generated and the propagated waves are always in phase. (d) Isotropic system with randomised orientation. The amplitude of the generated wave in each grain is random, leading to an overall randomised pattern. The phase shifts are identical to the ones in the system with fixed orientations, because the refractive indices are isotropic.}
    	\label{fig:illustration}
\end{figure}

    Fig.~\ref{fig:illustration}a-d visualizes how the waves are propagated and summed by the algorithm within the first three grains for four different configurations. For each configuration the top part of the image shows the $\bm{E}_{\text{gen}}$ and $\bm{E}_{\text{pre}}$ separately in each grain. The offsets from zero were added afterwards for visual clarity. The bottom part shows the sum of all the contributions $\bm{E}_{\text{tot}}$, which is equivalent to the total SH-field along the assembly.
    An isotropic system with a fixed orientation is shown in Fig.~\ref{fig:illustration}a. As expected, the generated waves are generally not in phase with the propagated waves from the previous crystals. This constant phase-mismatch leads to constructive and destructive interference in periodic intervals, leading to a $|\sin(x)|$ envelope function modulating the second harmonic E-filed along the structure. The maxima match odd and the minima even multiples of the coherence length.
    An isotropic system, with randomised grain orientations is shown in Fig.~\ref{fig:illustration}b. The amplitudes of the generated waves within each grain are randomised and the phases have arbitrary relations. Therefore, the total E-field amplitude will grow and decline in a random fashion. However, on average (considering multiple randomised sticks) the field amplitude grows with the square root of the number of domains. This ultimately leads to a linear dependence of the intensity on the number of domains, which is known as the RQPM.
    In a birefringent system with an not phase-matched fixed orientation for each grain, depicted in Fig.~\ref{fig:illustration}c, the fields are plotted in the lab-frame. Different colours are used for the components along the $\bm{a}$ and the $\bm{b}$ axis (dark/light blue). The SHG has different amplitudes and coherence lengths along the two different polarisation directions. In the propagated waves, the amplitudes of the two orthogonal beam components changes within one domain, because the birefringence introduces retardation, leading to an elliptically polarised beam which changes its eccentricity as well as its inclination during the propagation. Lastly, Fig.~\ref{fig:illustration}d shows a birefringent system in its phase-matched orientation, with $\vartheta = \vartheta_{\text{phm}}$, $\varphi = 0$, $\gamma = 0$, and the polarisation of the incoming beam is set to $\beta = 0$. This corresponds to pure Type-II ($oo$-$e$) phase matching. In this orientation, the generated and propagated waves are always in phase, which is correctly simulated by our algorithm. As a result, the amplitude of the total E-field increases linearly, which ultimately leads to the quadratic dependence of the intensity with the sample length. 

    \subsection{The Single Crystal Limit and Quasi-Phase Matching}
        \label{sec:simulation:reliability:boundary}
\begin{figure}[ht!]
        	\centering
        	\includegraphics[width=1.0\textwidth]{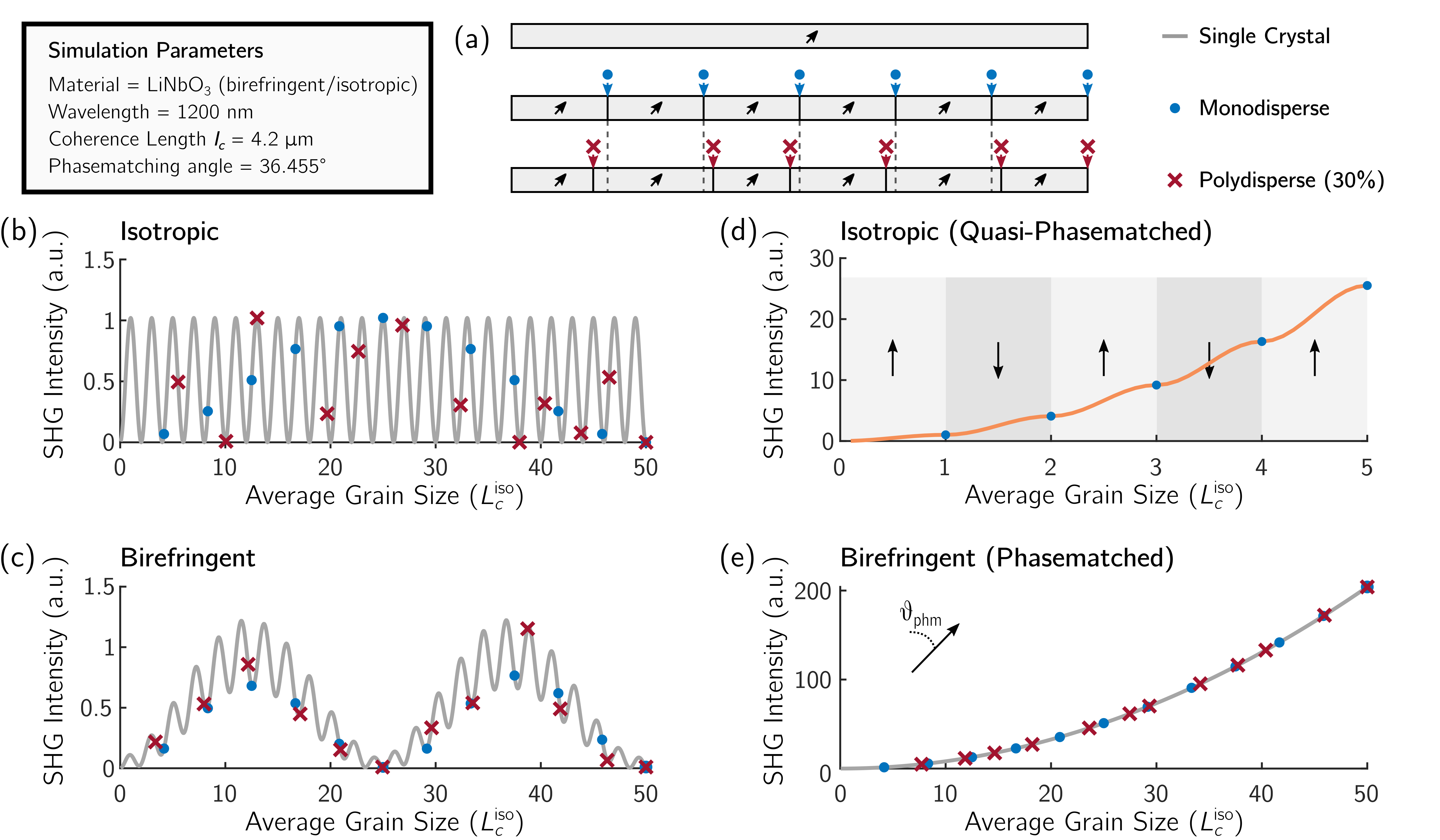}
        	\caption[Simulation: Comparison between the Single Crystal and Many Grains with the Same Orientation]{(a) Schematic representation a single grain split up in multiple grains of the same orientation either into monodisperse or polydisperse domains. In (b)-(e) the SHG intensity is normalised with the single grain peak intensity at $L_c^{\text{iso}} = 1.2\ \mu m$. (b) An isotropic assembly of 30 grains (size = $\frac{5}{3}$ \Lc) with a fixed orientation ( $\varphi=0$, $\vartheta=\pi/2$, $\gamma=\pi/2$) for every grain. (c) A birefringent assembly with 30 grains (size = $\frac{5}{3}$ \Lc) with a fixed orientation ($\varphi=0$, $\vartheta=\pi/2$, $\gamma=\pi/5$) for every grain. In \textbf{b} and \textbf{c} the intensity after each grain in the monodisperse assembly (blue dots) and in the assembly with 30\% polydispersity (red crosses) follows the behaviour in a single crystal of the same orientation (grey). (e) Simulation of the QPH in an isotropic medium, by periodic poling of the domains with a thickness of one coherence length. (f) Phase matching in a single crystal, polydisperse or monodisperse domains.}
        	\label{fig:sanity_checks}
        \end{figure}

        In order to verify that the amplitudes and phases are correctly propagated between the grains and that the dependence of the generated wave on the phase of the fundamental wave is implemented accurately, an array of grains with identical parameters is compared to the reference simulation of a single grain (Fig.~\ref{fig:sanity_checks}). The behaviour inside the single grain is directly derived from equation~(1) in the main text as a function of the length of the crystal, which serves as the reference of the physically expected behaviour since no phase considerations at grain interfaces are necessary for the calculation. This single grain reference is shown as a gray continuous line. The single grain is compared to an assembly of multiple grains adding up to the same total length, which all have the same orientation as the single crystal. The output received from the simulation at the end of each grain is depicted with blue dots for monodisperse domains and with red crosses for polydisperse domains, as illustrated in Fig.~\ref{fig:sanity_checks}a. This sanity check is performed for different systems with a well-known physical behaviour: Fig.~\ref{fig:sanity_checks}b,c,e show systems with a fixed crystal orientation for an isotropic crystal, a birefringent crystal and a phase-matched birefringent crystal. All cases match the single grain behaviour perfectly, confirming the correct grain to grain transmission and propagation of all involved waves. Interestingly, in the birefringent case, one can see the superposition of the different beam components with dissimilar coherence lengths. Lastly, we demonstrate that we can also simulate a quasi-phase matched system in Fig.~\ref{fig:sanity_checks}d. A quasi-phase-matched system consists of periodically poled layers with a thickness of one \Lc \ each~\cite{fejer1992QPM}. In our model, the poling is simulated by flipping the orientation of each grain with respect to the previous grain through a rotation of $\pi$ around the angle $\gamma$. The intensity plateaus at the end of the grains. Before it starts decreasing, the next grain begins (with a flipped orientation) and bringing the waves back in phase, thanks to a phase flip by $\pi$.
        
    \newpage
    \clearpage
    
    \section{SHG Intensity with Increasing Average Grain Size}
    
\begin{figure}[ht!]
        \centering
        \includegraphics[width=01.0\textwidth]{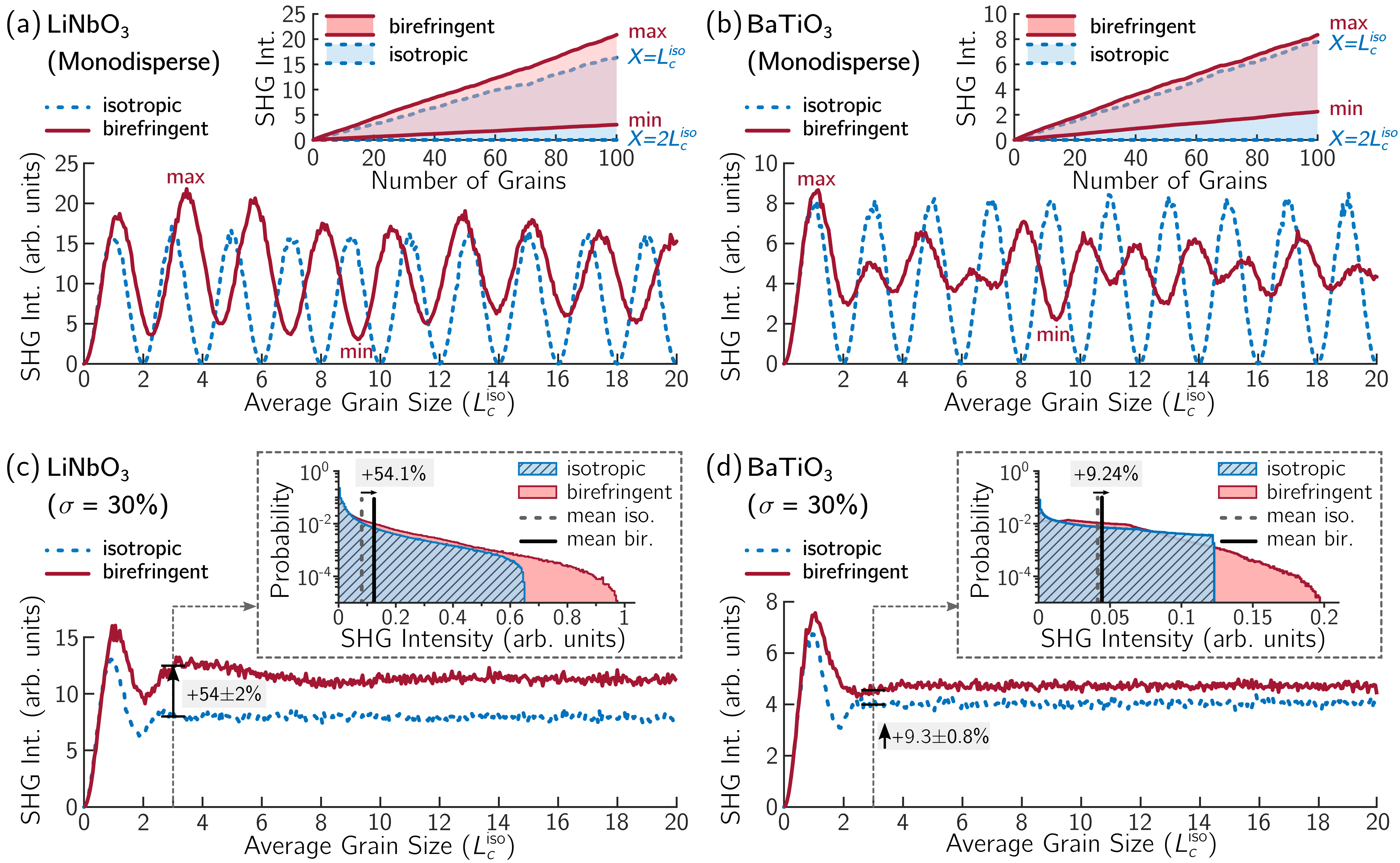}
        \caption{The SHG intensity of an assembly with 100 grains as a function of the average grain size $\bar{X}$, averaged over 1000 parallel sticks. (a),(b) In isotropic LiNbO$_3$ and BaTiO$_3$ (blue dotted) with monodispersed grain sizes, the SHG intensity shows maxima at odd and minima at even multiples of the coherence length. Including birefringence (solid red) the efficiency of the oscillation is offset from zero and the range of possible efficiencies is reduced. The insets of (a),(b) show the linear increase of the SHG intensity with the number of grains for different grain sizes. In LiNbO$_3$ the birefringent maximum (max) is at $X=3.47$ \Lc \ and the minimum (min) at $X=9.25$ \Lc. In BaTiO$_3$ the birefringent maximum is at one \Lc \ and the minimum at $X=9.2$ \Lc.
        (c),(d) With polydispersed grain sizes, the dependence on the average grain size disappears completely above 3 \Lc \ in isotropic materials. In the corresponding birefringent materials, the stable value is only reached for larger average grain sizes, at $\sim8$ \Lc \ in \LNO and at $\sim4$ \Lc \ in \BTO. The insets of (c),(d) show the distribution of the SHG intensity generated by the individual grains within the stick for isotropic and birefringent grains. The widening of the distribution leads to an increase of the average grain intensity. }
        \label{fig:disorder_funtion_of_grainsize}
    \end{figure}

    Fig.~\ref{fig:disorder_funtion_of_grainsize}a-b shows the SHG intensity scaling with increasing average grain size $0 \leq \bar{X} \leq 20$\Lc \ in monodispersed assemblies, both, isotropic and birefringent. In both materials, the minimal and maximal efficiency in the oscillation are increased due to the birefringence. In LiNbO$_3$ (Fig.~\ref{fig:disorder_funtion_of_grainsize}a), the efficiency increase is more pronounced, while in BaTiO$_3$ (Fig.~\ref{fig:disorder_funtion_of_grainsize}b) the extreme of the oscillation are narrowed after the initial peak at one \Lc. Interestingly, the local maxima and minima in the oscillation in both birefringent materials appear periodically. In LiNbO$_3$ this periodicity is larger than the \Lc, while in BaTiO$_3$ it is shorter than \Lc. In the insets of \ref{fig:disorder_funtion_of_grainsize}a-b, the linear increase of the SHG intensity with the number of grains is shown, for the grain sizes with the maximal and minimal efficiency (SHG intensity per grain). The shaded area represents all the possible values for different grain sizes. 
    
    Fig.~\ref{fig:disorder_funtion_of_grainsize}c-d shows the SHG intensity with increasing average grain size $0 \leq \bar{X} \leq 20$ \Lc \ in polydispersed assemblies. The results for LiNbO$_3$ (Fig.~\ref{fig:disorder_funtion_of_grainsize}c) are presented and discussed in the main text. The similar behaviour of BaTiO$_3$ is shown in Fig.~\ref{fig:disorder_funtion_of_grainsize}d. In contrast to LiNbO$_3$, BaTiO$_3$ does not show a second peak around 3 \Lc, but instead approaches directly a stable value for $\bar{X} > 3$ \Lc.
    
    \newpage
    \section{SHG Intensity Distribution of Single Grains Smaller than the Coherence Length}
    \label{sec:efficiency} 
\begin{figure}[ht!]
        \includegraphics[width=1.0\textwidth]{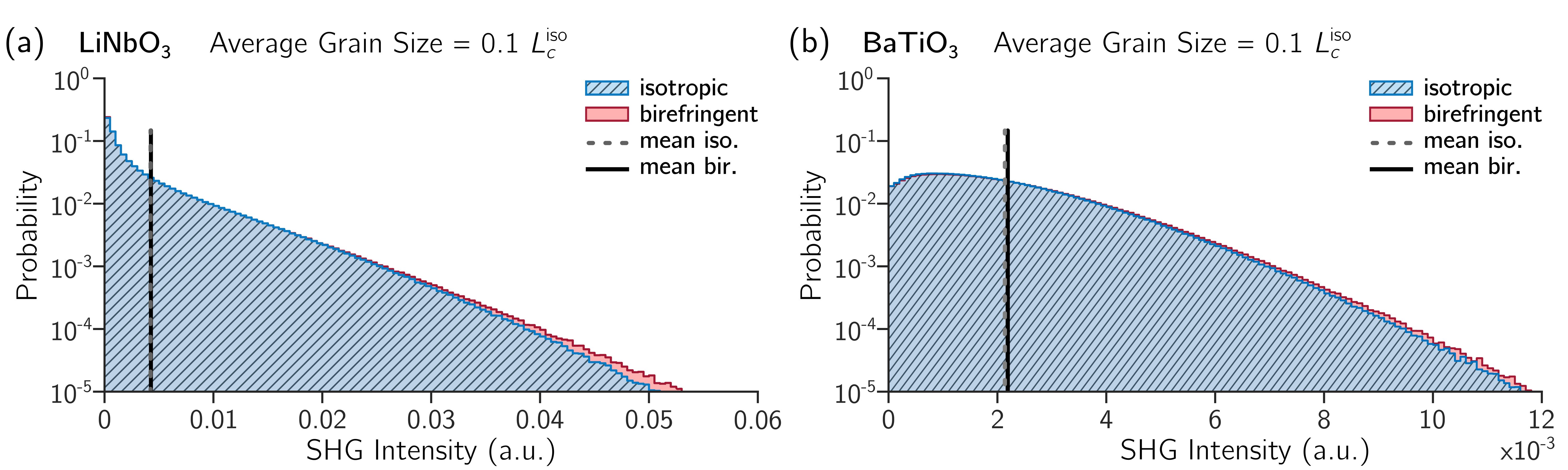}
        \caption{Distribution of the SHG intensities for $10^5$ randomly oriented single grains, with an average grain size $\bar{X} = 0.1$ \Lc \ and with $\sigma = 30\%$ polydispersity for LiNbO$_3$ (a) and  BaTiO$_3$ (b). The mean intensities of all grains (vertical lines) are identical for the isotropic and the birefringent case for both materials.}
        \label{fig:efficiency_Distribution}
    \end{figure}

    The increase of the mean SHG intensity of the individual grains inside the simulated stick due to the birefringence of the grains is responsible for the efficiency increase of the birefringent assembly in comparison to its isotropic analogue. However, this only applies for grains larger than the isotropic coherence length \Lc, where the increase of the orientation specific coherence length \Lcbir \ in the birefringent medium provides an advantage. In Fig.~\ref{fig:efficiency_Distribution}, one can in fact see, that for grains smaller than \Lc \ (average grain size $\bar{X} = 0.1$ \Lc) the SHG distributions of the individual grains in the stick is almost identical for both, isotropic and birefringent grains, independent of the material. In this small grain size regime, the maximal efficiency is limited by the size of the grain $X_n$, rather than its coherence length \Lcbir. We note that there is still a slight shift in the distribution favouring the birefringent material. Interestingly, we also observe that the distribution of the single grain intensities differs significantly for the two materials.

\newpage
\cleardoublepage
    
\section{SHG Intensity Scaling for Phase-matchable Lithium Niobate}

\begin{figure}[ht!]
    \centering
        \includegraphics{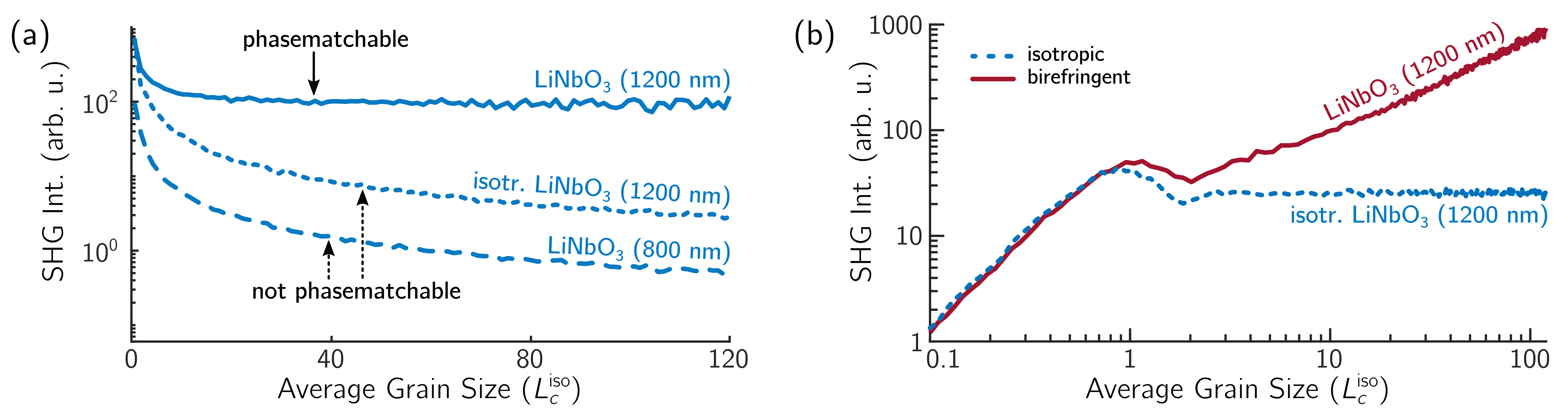}
        \caption{(a) The SHG intensity scaling with the average grain size for an assembly with a constant thickness. The LiNbO$_3$ at 1200 nm is phase-matchable (solid line) and displays a constant SHG intensity for large average grain sizes. The isotropic LiNbO$_3$ at 1200 nm as well as the birefringent LiNbO$_3$ at 800 nm are not phase-matchable and show the same $1/X$ scaling with the average grain size. (b) SHG intensity scaling with the average grain size for a LiNbO$_3$ assembly with a constant number of grains at a phase-matchable wavelength of 1200 nm.}
        \label{fig:phasematchability}
    \end{figure}

The scaling of the SHG intensity of an assembly depends on the phase-matchability of the grains. In the main manuscript, we used the isotropic analogue as the not phasematchable reference. However, instead of using the isotropic analogue, not phase-matchable LiNbO$_3$ can be obtained at a wavelength of 800 nm, as shown in section \ref{sec:simulation:theory:phasemismatch}. For this case, Fig.~\ref{fig:phasematchability}a shows the scaling of the SHG intensity with the average grain size. While the LiNbO$_3$ at 1200 nm, which is phasematchable, shows a constant SHG scaling. Both the isotropic analogue at 1200 nm and the LiNbO$_3$ at 800 nm show the same scaling behaviour. The only difference is a shift in the overall intensity due to the different wavelength.

In the main manuscript we showed the scaling of the SHG intensity in a phase-matchable situation for samples with a constant thickness (constant stick length). In Fig.~\ref{fig:phasematchability}b, we show the scaling of the SHG intensity in birefringent grains (phase-matchable) and isotropic grains with a constant number of grains, instead of a constant sample thickness. The birefringence introduces an enhancement of the SHG intensity, which grows linearly with the average grain size bringing the birefringent phasematchable grains into a new scaling regime. This linear scaling is consistent with the results found by Aramburu et al.~\cite{Aramburu2013Second}.

\bibliography{supplementaries}